\documentclass[12pt,fleqn]{article}
\usepackage{amsmath}
\usepackage{slashed}
\usepackage{url}
\usepackage{bm}
\usepackage{graphicx}
\usepackage[utf8]{inputenc}
\usepackage[english]{babel}
\usepackage{amsmath}
\usepackage{amsfonts}
\usepackage{amssymb}
\usepackage{graphicx}
\usepackage{float}
\usepackage{hyperref}
\begin{document}

\title{Interaction of a black hole with scalar field in cosmology background}
\author{Malik Almatwi\footnote{malik.almatwi@gmail.com}\,\, }

\date{Department of Theoretical Physics, Faculty of Science, University of Mazandaran, 47416-95447, Babolsar, Iran}
\maketitle

\tableofcontents

\section{Abstract}
Recent data from elliptical galaxies indicate that the growth in the masses of black holes exceeds what is expected from the accretion of surrounding matter, and this growth appears to be dependent on the expansion of the universe. This phenomenon can be explained by considering the accretion of dark energy, which is responsible for cosmological expansion, into these black holes. In this paper, we investigate the perturbative interaction of a black hole with a real scalar field $\phi$ (which can represent dark energy) in a cosmological background using an appropriate metric. We derive solutions for the field $\phi(t, r)$, the black hole mass $M(t)$, and the expansion rate $H(t)$, and discuss the behaviour of the scalar field $\phi$ in the vicinity of the black hole, with respect to exterior and interior observers. We obtain the energy density of $\phi$ inside of the black hole and find that energy density converges and tends to take a non-vanishing fixed values (stable) at late time of cosmological expansion. We find that on horizon surfaces, it is possible to make the field $\phi$ continuous and bounded (maybe differentiable) function.
\\
\\
{\bf Key words}: Real scalar field;  Black hole;  Expansion of the universe; Accelerated expansion; Black hole mass; Interior solutions.

\section{Introduction}
The real scalar field provides a dynamical representation of the universal dark energy, its crucial role is providing a universal negative pressure, so explaining the accelerated expansion of the universe \cite{Copeland}, describing the inflaton as the driver of the early time cosmological inflation  \cite{Baumann} and dark energy as responsible for late time cosmic speed up \cite{Bamba, Amendola}, in addition to the possibility of using any potential function for the real scalar field and constructing additional interactions such as non-minimal couplings \cite{Ishwaree, Mostaghel, Moraes, Nikodem, Bertolami}. In order to respect the cosmological principles (universe is both homogenous and isotropic at large distances scales), some suitable geometries such as Schwarzschild-de Sitter and McVittie geometries where constructed, and so by solving Einstein equation one can find the behaviour of the black holes under influences of the cosmological expansion \cite{Brien, Nemanja, Gregoris, McVittie, Gabbanelli, ValerioFaraoni}. 
\\
\\

Astronomically, for large scales of times and distances, an experimental data obtained from elliptical galaxies indicates that there is a cosmological growth in masses of the black holes more than that the expected by the accretion of surrounding matter, and more massive black holes tend to reside in more massive galaxies, and the growth in black holes masses depends on the redshift, with a higher factor at higher redshifts, obeying the formula; $M(a)= M(a_i)(a/a_i)^k$, where $a_i$ is the RW scale factor at which the black hole becomes cosmologically coupled, and $k > 0$ is cosmological coupling parameter. Therefore, the black holes can be regarded as non-singular cosmological objects. That cosmological coupling leads to an increasing in masses of the black holes with the expansion of the universe independently of accretion or mergers. That behaviour can be tested by considering the growth of supermassive black holes in elliptical galaxies over the redshifts $0 < z \le 2.5$. Experientially, $k\approx 3$, which makes the internal energy density of the coupled black holes approximately constant \cite{Duncan, Duncan1, Sohan, Valerio, Kevin, Hengxin}.
\\

The cosmological growth in the masses of the coupled black holes during the cosmological expansion, with keeping their interior energy densities approximately constant (same feature of vacuum energy), led to believe in existence of a linking between that coupled black holes and the dark energy, such as regarding them as sources of the dark energy, i.e, after the gravitational collapse of baryonic matter, that matter could convert to dark energy that resides interior of that black holes without singularities, and that dark energy drives the expansion of the universe. Other viewpoint is that suggesting an accretion of the universal dark energy into the black holes (due to the gravitational force), leading to a cosmological growth in the masses of that black holes, and finally, the black holes become repositories of the dark energy. In that model, the dark energy is considered as a perfect fluid that fills the whole universe satisfying $\rho+p\ge 0$ \cite{Lobo}-\cite{Aaron}. 
\\

In our work, we study spherical symmetric gravitational interaction of a black hole with a real scalar field $\phi(t,r)$ (which can represent the dark energy) by considering the local behaviour of that field (by its dependence on $r$). That locality of the field $\phi(t,r)$ relates to the gravitational interaction with the black hole and determines the possibility of concentrating (or creating) that field inside the black hole or around it. But as we will see, there are also solutions that give a global aspects for $\phi$ (by its independence of $r$), and other solutions relate to ground state energy of the field $\phi$, that fills the universe homogeneously representing a vacuum energy (i.e, cosmological constant). Therefore, if we combine (by sum) that two solutions, we obtain a total solution for $\phi(t,r)$ that can describe both local and global aspects of the dark energy. Therefore in our work, the assumption that the dark energy is a perfect fluid is not needed; actually, a scalar field with any potential and for any solution can not be described by a perfect fluid, i.e, the equation of state $p=p(\rho)$ is not necessary coordinates independent. And the representation of the dark energy by a perfect fluid is not enough to explain its attractive interaction with the black holes, particularly, the dark energy provides a repulsive force, therefore we need to represent it by a canonical real scalar field (Quintessence), and since that field is canonical, it can explain the reality of existence of attractive force between the black holes and the dark energy, and whether that attractive force survives during the process of dark energy accretion into the black holes. In our work, we find that there are solutions for the scalar field that are concentrated around the singularity of the space-time, in other words, that solutions are concentrated inside the black hole. We can also regard that behaviour as a creation of the real scalar field inside the black hole. This gives a possibility for regarding the black holes as accretion of the dark energy. We obtain a formula for the mass function $M(t)$ that is approximately the same as $M(a)= M(a_i)(a/a_i)^k$, and find that the masses of the black holes are always in increasing during the cosmological expansion, but in our work, we use only a black hole and a real scalar field.  
\\
\\

To study the gravitational interaction of a black hole (without spin) with a real scalar field $\phi(t,r)$ in cosmology background, we suggest some metric that can describe a black hole of a time dependent mass $M(t)$, and that black hole is immersed in an expanding universe with a time dependent expansion rate $H(t)$, in addition to including an arbitrary function $F(t,r)$. We will find that the function $F(t,r)$ contributes in spherical symmetric distribution of the kinetic and potential energies of the field $\phi(t,r)$ at each moment of time. By that, our system includes four unknown functions $M(t)$, $H(t)$, $F(t,r)$ and a real scalar field $\phi(t,r)$. We derive Einstein equation of that system and obtain a differential equation for $F(t,r)$, and suggest some solutions regarding the fact that at late times of the expansion of the universe, the energy density of the field $\phi$ decreases due to the cosmological expansion, therefore the function $F(t,r)$ also decreases. We study two solutions of $F(t,r)$, a solution with decreasing with powers of $1/t$ (slow expansion) and other with powers of $\exp(-t)$ (fast expansion). We find that the second solution can give an accelerated expansion of the universe, but with non-vanishing ground state values of the potential of $\phi$. We determine the potential $V(\phi)$ of $\phi$ in terms of the function $F(t,r)$, and obtain the solutions of the expansion rate $H(t)$, the solutions of the black hole mass $M(t)$ and find that the black hole mass increases during the expansion of the universe, but slightly, until taking a constant value, therefore the black holes are stable in view of the expansion of the universe. We obtain also the solutions of $\phi(t,r)$ inside the black hole by using the swapping $t\to r$ and  $r\to t$, and obtain the energy density of $\phi(t,r)$ inside the black hole (as measured by some interior observer).
\\

\section{Suggesting a metric and discussion the solutions}
In this section, we derive Einstein equation of a real scalar field using some metric and obtain some possible approximated solutions of the differential equations at late time of expansion of the universe. In order to study a black hole of time dependent mass $M(t)>0$, that is immersed in an expanding universe, with a time dependent expansion rate $H (t)>0$, we use the metric
\begin{equation}\label{eq:1}
\begin{split}
{ds}^{2}|_{r\ge R_{BH}}=g_{\mu \nu} dx^\mu dx^\nu&=-\left(1-\frac{2 M (t)}{r}-r^{2} H (t)^{2} \right) F (t, r)^{2}\mathit{dt}^{2}\\
&+\frac{1}{1-\frac{2 M (t)}{r}-r^{2} H (t)^{2}}\mathit{dr}^{2}+r^{2} \left( d\theta^2 + \sin^2(\theta) d \Phi ^2 \right) \, ,
 \end{split}
\end{equation}
and
\begin{equation}\label{eq:a69}
\begin{split}
{ds}^{2}|_{r<R_{BH}}=\frac{1}{1-\frac{2 M (r)}{t}-t^{2} H (r)^{2}}  &  \mathit{dt}^{2} -\left(1-\frac{2 M (r)}{t}-t^{2} H (r)^{2} \right) F (r, t)^{2}\mathit{dr}^{2}\\
&+t^{2} \left( d\theta^2 + \sin^2(\theta) d \Phi ^2 \right) \, ; \, 0<t<R_{BH} \, .
 \end{split}
\end{equation}
This metric describes an expanding universe including a black hole in $r=0$ (the centre of its mass). An observer who uses this metric will note that the universe expands in all directions with spherical symmetry and measures two radial forces; an attractive force, in opposite direction of $\vec r$, comes from the gravity of the black hole, and other comes from the expansion of the universe in direction of $\vec r$. That expansion can be obtained by using a real scalar field which is needed for providing a universal negative pressure, and that pressure is needed for obtaining an expansion force against the gravitational attractive force of the universal matter and energy. In our study, the real scalar field $\phi$ is a function of both radial coordinate $r$ and time $t$, with spherical symmetry. Actually, the locality of $\phi$ comes from the attractive force of gravity of the black hole. As we will see, according to behaviour of the solution of $\phi$, that field is concentrated (or created) inside the black hole (mostly in $r\approx 0$). The function $F(t,r)$ has to be differentiable for all $t, r\ne 0$ and does not have any special behaviour on horizon surfaces. 
\\

At large distances, $M (t)/{r}<<1$, the metric (\ref{eq:1}) limits to RW metric, in which, $H (t)$ is defined by $H=\dot a/a$, where $a(t)$ is universal scale factor, therefore the effect of the black hole at high distances is neglectable. While in small distances, $r^{2} H (t)^{2}<<1$, it limits to Schwarzschild metric and so the effect of the expansion of the universe in this case is neglectable. This metric gives two Horizon surfaces \cite{ValerioFaraoni}, given by the positive solutions of $(g_{rr})^{-1}=0$, therefore it is a appropriate for studying a black hole in an expanding universe. At late times of the cosmological expansion, the universal critical energy $3 H (t)^{2}$ (as defined in Friedman equation) becomes small enough so that the cosmological horizon radius $R_C (t)$ becomes large enough comparing with the black hole horizon radius $R_{BH} (t)$, so we get $R_{BH} /R_C <<1$ and $R_C \approx H^{-1} >> R_{BH}\approx 2M$. Actually, these approximations allow us to get the solutions easily.
\\

In this system, we have four unknown functions; $F(t, r)$, $M(t)$, $H (t)$ and $\phi (t,r)$. The Einstein equation of the metric (\ref{eq:1}) gives four equations, therefore the number of unknown functions equals to the number of equations. The potential $V(\phi )$ of $\phi (t,r)$ is assumed to be given, so it is considered as a known function, however we will find that the solution $f_0 (r)$ of the function $F(t,r)$ relates with the potential $V(\phi )$ in some equation. 
\\

We start by deriving the Einstein equation, then search for the possible solutions. The corresponding Einstein tensor of the metric (\ref{eq:1}) is
\begin{equation}\label{eq:2}
G_{tt}  =  - \frac{{3\left( {r^3 H(t)^2  + 2M(t) - r} \right)}}{r}H(t)^2 F(t, r)^2  =  - 3H(t)^2 g_{tt} ,\\
\end{equation}
\[
G_{tr}  =  - \frac{{2\left( {r^3 H(t)\dot H(t) + \dot M(t)} \right)}}{{r\left( {r^3 H(t)^2  + 2M(t) - r} \right)}},   \\
\]
\begin{equation*}
\begin{split}
&G_{rr}  = \frac{{3rH(t)^2 }}{{r^3 H(t)^2  + 2M(t) - r}}+\frac{{2\left( {r^3 H(t)^2  - r + 2M(t)} \right) {F'(t, r)} }}{{rF(t, r)(r^3 H(t)^2  + 2M(t) - r)}} \\
 &\quad=  - 3H(t)^2 g_{rr}  - \frac{2F'(t, r)}{rF(t, r)}Kg_{rr},
 \end{split}
\end{equation*}
and
\begin{equation*}
\begin{split}
& G_{\theta \theta }  =- 3r^2 H(t)^2  - \frac{{r^2 F''(t, r)}}{{F(r)}} K - \left( {4rK - 9M(t) + 3r} \right)\frac{F'(t, r)}{{F(t, r)}}  \\ 
 & - \frac{1}{{ {K}^3 F(t, r)^2 }}\left\{ {r K{\frac{d}{{dt}}\left( {r^3 H(t)\dot H(t) + \dot M(t)} \right)}  - 4\left( {r^3 H(t)\dot H(t) + \dot M(t)} \right)^2  } \right\} \, ,\\ 
&\\
&+ \frac{1}{{K^3 F(t, r)^2 }}\left[ {rKF(t, r)^{ - 1} \dot F(t, r) \left( {r^3 H(t)\dot H(t) + \dot M(t)} \right)} \right] \, , \\
&\\
&G_{\Phi\Phi}=G_{\theta\theta} \sin^2(\theta) \, .
 \end{split}
\end{equation*}
We have used
\[
K(t,r) = r^2 H(t)^2  +\frac{2M(t)}{r}  - 1,     \quad     \dot H(t) =\frac{d H (t)}{dt} \quad \text{and}  \quad F'(t, r)=\frac{\partial F(t, r)}{ \partial r} \, .
\]
We note that $(g_{rr}(t, r))^{-1} =-K$, therefore $K=0$ is equation of horizon surfaces ($r=R_H (t)$). We also note that the term $r^3 H(t)\dot H(t) + \dot M(t) $ is included only in equations of $G_{tr  }$ and $G_{\theta \theta }$, therefore these two equations give the equations of $\dot H(t)$ and $ \dot M(t)$. 
\\

The Lagrangian of a canonical real scalar field $\phi $, with respect to a metric $g$, is
\[
L(\partial _\mu  \phi ,\phi ,g) = - \frac{1}{2}g^{\mu \nu } \partial _\mu  \phi \partial _\nu  \phi  - V\left( \phi \right) \, .
\]
And its energy-momentum tensor is given by
\begin{equation}\label{eq:a67}
T_{\mu \nu } ^{(\phi)} = \partial _\mu  \phi \partial _\nu  \phi  + g_{\mu \nu } L^{\left( \phi \right) }= \partial _\mu  \phi \partial _\nu  \phi  + g_{\mu \nu } \left( { - \frac{1}{2}g^{\rho \sigma } \partial _\rho  \phi \partial _\sigma  \phi  - V \left( \phi \right) } \right) \, .
\end{equation}
The corresponding Einstein equation $G_{\mu \nu }  = 8\pi T_{\mu \nu } ^{(\phi)}$, with using the metric (\ref{eq:1}), implies (with using units of $G_N=1$)
\begin{equation}\label{eq:3}
\begin{split}
 &G_{tt}  =  - 3H(t)^2 g_{tt}   = 8\pi \left( {\frac{1}{2}\dot \phi ^2  - \frac{1}{2}g_{tt} g^{rr} \phi{ '}^2  - g_{tt} V\left( \phi  \right)} \right) \, , \\
&\\
&G_{tr}  =  - \frac{{2\left( {r^3 H(t)\dot H(t) + \dot M(t)} \right)}}{{r\left( {r^3 H(t)^2  + 2M(t) - r} \right)}}=8\pi \dot \phi \phi ' \,  ,   \\
&\\
& G_{rr}  =  - 3H(t)^2 g_{rr}  + \frac{{2F'(t, r)}}{{rF(t, r)}} = 8\pi \left( {\frac{1}{2}\phi {'}^2  - \frac{1}{2}g_{rr} g^{tt} \dot \phi ^2  - g_{rr} V\left( \phi  \right)} \right)  \, ,\\ 
 \end{split}
\end{equation}
and
\begin{equation}\label{eq:4}
\begin{split}
 G_{\theta \theta }  &=  - 3r^2 H(t)^2  - \frac{{r^2 F''(t, r)}}{{F(t, r)}}K - \left( {4rK - 9M(t) + 3r} \right)\frac{{F'(t, r)}}{{F(t,r)}} \\ 
&  - \frac{1}{{K^3 F(t,r)^2 }}\left\{ {rK {\frac{d}{{dt}}\left( {r^3 H(t)\dot H(t) + \dot M(t)} \right)}  - 4\left( {r^3 H(t)\dot H(t) + \dot M(t)} \right)^2 } \right\} \,  \\ 
& + \frac{1}{{K^3 F(t, r)^2 }}\left[ {rKF(t, r)^{ - 1} \dot F(t, r) \left( {r^3 H(t)\dot H(t) + \dot M(t)} \right)} \right] \,  \\
&\\
 & = 8\pi T_{\theta \theta }  = 8\pi g_{\theta \theta } L^{\left( \phi  \right)} \, . \\ 
\end{split}
\end{equation}
By removing the potential $V(\phi )$ from equations of $G_{tt}$ and $G_{rr}$, we simply obtain
\[
g_{tt} \left( {g^{tt} G_{tt}  - g^{rr} G_{rr} } \right) = 8\pi \left( {\dot \phi ^2  - g_{tt} g^{rr} \phi{'}^2 } \right) = 8\pi \left( {\dot \phi ^2  + K^2 F^2 {\phi '}^2 } \right) \, ,
\]
which becomes
\begin{equation}\label{eq:25}
 \frac{{FF^{\prime}}}{r}K^2  =8\pi \left(\frac{1}{2} \dot \phi ^2  +\frac{1}{2} K^2 F^2 \phi {'}^2  \right) \, .
\end{equation}
This equation implies that the function $F(t,r)$ has to satisfy $FF^{'}>0$, for all values of $r>0$. This equation is well defined on horizon surfaces ($K=0$), but with $\dot \phi =0$, while $\phi {'}$ does not necessary vanish. The term $ (\dot \phi ^2  + K^2 F^2 {\phi'}^2)/2$ can be considered as a kinetic energy of $\phi$.
\\

And by omitting the term $\dot \phi ^2  + K^2 F^2 \phi{ '} ^2$ from equations of $G_{tt}$ and $G_{rr}$, we obtain the value of the potential in terms of the metric functions, we get
\begin{equation}\label{eq:17}
8\pi V\left( \phi  \right)   =3H(t)^2+  \frac{{ {F'(t,r)}  }}{{rF(t,r)}} K\, .
\end{equation}
Actually, the potential $V\left( \phi  \right)\ge0$ depends only on $\phi$, not on the coordinates $(t, r, \theta, \Phi) $, but for a solution, like $\phi(t, r, \theta, \Phi)$, then $\left. {8\pi V\left( \phi  \right)} \right|_{\phi (t,r,\theta ,\Phi )} $ is just assigning a value to the potential $V\left( \phi  \right)$ in a point  $(t, r, \theta, \Phi) $. 
\\

Regarding the equations (\ref{eq:25}) and (\ref{eq:17}), we see that the kinetic and potential energies of $\phi$ are well defined for all $(t, r)$; $r\ne 0$, including horizon surfaces, $r=R_H$ ($K=0$), although the metric (\ref{eq:1})  is not defined on that surfaces, and we can also make $\phi$ continuous (and bounded) on that surfaces (in this work, we do not do that). The remaining problem is with the equation $G_{tr}=8\pi \dot \phi  \phi '$,  but we will find (in the Appendix) that the term $r^3 H(t)\dot H(t) + \dot M(t)$ is proportional to $K^2$, this implies also vanishing of $\dot \phi $ on horizon surfaces, so regarding the field $\phi$ as a continuous (and bounded) on that surfaces is possible. 
\\

Therefore, the field $\phi (t, r)$ can be defined on horizon surfaces ($K=0$) as a continuous bounded function (maybe differentiable), where $F(t,r)$ is a non-vanishing differentiable bounded function for all $r\ne 0$. In other words, the field $\phi$ itself does not feel the horizon event, but when we quantize it to get particles, those particles will be effected by the horizon event. So, may we think that the horizon event is a problem with the quantization of the fields, but not a problem with the fields themselves. We can glue the exterior and interior solutions of $\phi(t,r)$ on the black hole horizon surface ($r=R_{BH}$) by imposing $\phi_{in}(t,r)|_{r=R_{BH}}=\phi_{ex}(t,r)|_{r=R_{BH}}$ and $\phi{'}_{in}(t,r)|_{r=R_{BH}}=\phi{'}_{ex}(t,r)|_{r=R_{BH}}$, the last condition implies $\phi{'}_{ex}(t,r)|_{r=R_{BH}}=0$ (because of $\phi{'}_{in}(t,r) |_{r=R_{BH}} =0$). But this glueing gives other formulas for $H(t)$ and $M(t)$ unlike what we will get in this work, therefore we do not do that gluing.
\\

It seems that there is contradiction on $K=0$ surfaces, when $8\pi V( \phi |_{K=0} )   =3H(t)^2 \ne \text{constant}$, because of $\dot \phi |_{K=0}    =0$. But
\[
d \phi |_{K=0}    =\dot \phi dt |_{K=0}  + \phi {'} dr |_{K=0} =\phi {'} dr |_{K=0}  \ne 0 \, , \quad \text{when}\quad  \phi {'} |_{K=0}    \ne 0 \, ,
\]
which implies that $\phi  |_{K=0}    \ne \text{constant}$, so $8\pi V( \phi |_{K=0} )   \ne \text{constant}$, as required for getting $H(t) \ne \text{constant}$. But it must be $ \phi {'} |_{K=0}    \ne 0$ always satisfied.
\\

It is clear that on the horizon surfaces, $K=0$ ($r=R_H (t)$), the potential (\ref{eq:17}) takes its highest value, $3H(t)^2$, which is the universal critical energy density. This is acceptable since, as we will see, the kinetic energy of $\phi$ vanishes on that surfaces. Therefore, $3H(t)^2$ is still regarded as a critical energy density and it is the highest possible energy density that the field $\phi$ can take. A same thing we see in the case of the spatially homogenous isotropic space (RW space-time) without black holes.
\\

Regarding the previous equations, we conclude that the function $F(t, r)$ has some physical role, that is, it contributes in spherical symmetric distributions of the kinetic energy (\ref{eq:25}) and the potential energy (\ref{eq:17}) over the space $(0, \infty) \times S^2$ at each moment of time. And, the concentration of $F(t, r)$ around the point $r=0$ implies a concentration of the field $\phi(t, r)$ around the same point. In general (as we will see), the behaviour of $F(t, r)$ determines the behaviour of $\phi (t, r)$.
\\

The metric (\ref{eq:a69}) describes the interior of the black hole ($r<R_{BH}$). It is clear that we can obtain the metric (\ref{eq:a69}) from the metric (\ref{eq:1}), just by the swapping $t$ to $r$ and  $r$ to $t$. Actually, in $r<R_{BH}$, the $g_{tt}$ component of the metric (\ref{eq:1}) becomes positive ($g_{tt} > 0$) and its $g_{rr}$ component becomes negative ($g_{rr} < 0$). Therefore, in order to preserve the future direction (defined by $-ds^2 >0$, $dt >0$) inside the black hole, the interior observers need to swap the coordinates; $t\to r$ and $r\to t$ in the metric (\ref{eq:1}), so swapping its components as; $g_{tt} (t, r) \to g_{rr} (r, t)$ and $g_{rr} (t, r) \to g_{tt} (r, t)$, so obtaining the metric (\ref{eq:a69}). 
\\

Therefore, this swapping does not change the coordinates $(t,r)$, but the problem is that when the time-like trajectories (coming from out of the black hole) cross the horizon surface of the black hole, they turn to space-like trajectories (as seen by exterior observers who use the metric (\ref{eq:1})), that means that the dynamics stops and the time loses its role, and the energy becomes like as potential energy (as the Coulomb potential in electromagnetic); Each particle became constrained (enforced) on a point on a space-like trajectory because of the strong gravity inside of the black hole, and so it does not exchange energy (no causal correlations) with others (as seen by exterior observers). In order to recover the dynamics, the interior observers need to change the metric (\ref{eq:1}), in a way that the space-like trajectories inside of the black hole turn to time-like trajectories and the dynamics is recovered again by using the metric (\ref{eq:a69}). That can be explained by the fact that the exterior observers regard the black hole as increasing of the gravity without limit, so the top effect occurs by turning the time-like trajectories to space-like trajectories (stop dynamics).
\\

Actually, this swapping does not change the space-time manifold, but it is just a re-labelling of the coordinates \cite{Rosa}, to obtain the metric (\ref{eq:a69}). That re-labelling is possible because the equations of $\phi(t,r)$, with respect to the metric (\ref{eq:1}), are defined for all $r\ne0$, and since the Einstein tensor depends only on the metric, so that re-labelling is right and implies also a swapping in the solutions of $\phi(t,r)$ to $\phi(r,t)$, where as we will see, the equations of $\phi(t,r)$ depend only on the used metric. 
\\

By dividing equation (\ref{eq:25}) by $G_{tr}  =8\pi \dot \phi \phi ^{\prime}\ne 0$, we obtain
\begin{equation}\label{eq:a70}
\frac{{2FF^{\prime}}}{{rG_{tr} }}K^2  = \frac{{\dot \phi }}{{\phi ^{\prime}}} + K^2 F^2 \frac{{\phi^{\prime}}}{{\dot \phi }} \, ,
\end{equation}
which implies that 
\[
K^2 F^2 \left( {\frac{{\phi^{\prime}}}{{\dot \phi }}} \right)^2  - \frac{{2FF^{\prime}}}{{rG_{tr} }}K^2 \left( {\frac{{\phi^{\prime}}}{{\dot \phi }}} \right) + 1 = 0 \, .
\]
This equation has the solutions
\[
\frac{{\phi{'}}}{{\dot \phi }} = \frac{{\frac{{2FF^{\prime}}}{{rG_{tr} }}K^2 + a \sqrt {\left( {\frac{{2FF^{\prime} }}{{rG_{tr} }}K^2 } \right)^2  - 4K^2 F^2 } }}{{2K^2 F^2 }}   \,,
\]
for $a=\mp 1$. 
\\

Multiplying the last equation with the equation $8\pi \dot \phi \phi ^{\prime}=G_{tr}  $, we obtain
\begin{equation}\label{eq:6}
\begin{split}
 8\pi \phi{ '}  ^2  &= G_{tr} \frac{{\frac{{2FF'  }}{{rG_{tr} }}K^2  +a \sqrt {\left( {\frac{{2FF'  }}{{rG_{tr} }}K^2 } \right)^2  - 4K^2 F^2 } }}{{2K^2 F^2 }} \\
 &=\frac{{F'  }}{{rF}} +a \sqrt {\left( {\frac{{F'  }}{{rF}}} \right)^2  - \frac{{G_{tr}^2 }}{{K^2 F^2 }}} \, ,
\end{split}
\end{equation}
and by using it back in equation (\ref{eq:25}), we get
\begin{equation}\label{eq:14}
\begin{split}
 8\pi \dot \phi ^2 = \frac{{FF'  }}{r}K^2  -a K^2 F^2 \sqrt {\left( {\frac{{F'  }}{{rF}}} \right)^2  - \frac{{G_{tr}^2 }}{{K^2 F^2 }}} \, .
\end{split}
\end{equation}
These two equations determine $ \phi '$ and $\dot \phi$ in terms of the metric functions; $H(t)$, $M(t)$ and $F(t,r)$, therefore obtaining the solutions of $ \phi$ in terms of that functions. But we need some enough equations to obtain solutions of the metric functions. As mentioned under the equation (\ref{eq:17}), the real scalar field $ \phi $ can be defined on horizon surfaces ($K=0$) as a continuous bounded function (maybe differentiable), and since $F(t,r)$ is a differentiable non-vanishing bounded function for all $r\ne 0$, that implies a boundedness for ${{G_{tr} }}/{{K F }}$ nearby horizon surfaces; by getting closer to $r=R_H$, in order to maintain the reality of $ \phi $, it must be $F'  /rF - G_{tr} /K F \ge 0$ always satisfied. As we will see, this issue will be simply solved by vanishing the root term.
\\

We note that the root term includes the function $G_{tr}$ which is given by the second equation of (\ref{eq:3}). But $G_{tr}$ includes the term $r^3 H(t)\dot H(t) + \dot M(t)$. We also note that the equations $G_{tr}= 8\pi T^{(\phi)}_{tr}$ and $G_{\theta \theta}= 8\pi T^{(\phi)}_{\theta \theta}$ are the only equations that include the term $r^3 H(t)\dot H(t) + \dot M(t)$, therefore we can get it from those two equations. 
\\

But in order to handle the equation $G_{\theta \theta}= 8\pi T^{(\phi)}_{\theta \theta}=8\pi g_{\theta \theta} L^{(\phi)}$, equation (\ref{eq:4}), we need to find the formula of the Lagrangian $ L^{(\phi)}$ in terms of the metric functions. We can get that Lagrangian with its simplest form by using the equation $G_{rr}=8\pi T_{rr } $; we obtain
\begin{equation}\label{eq:5}
\begin{split}
 G_{rr}  = 8\pi \left( {\phi '^2  + g_{rr} L^{\left( \phi  \right)} } \right) \Rightarrow 8\pi L^{\left( \phi  \right)} & = g^{rr} G_{rr}  - 8\pi g^{rr} \phi '^2  \\ 
  &= g^{rr} G_{rr}  + 8\pi K\phi '^2  \, . 
\end{split}
\end{equation}
And by using the formula of $G_{rr}$ and the solution of $\phi '^2$ (equation (\ref{eq:6})), we obtain
\begin{equation}\label{eq:7}
8\pi\left. {L^{\left( \phi  \right)} } \right|_{\phi \left( {t,r} \right)} =  - 3H\left( t \right)^2  - \frac{{F^\prime  }}{{rF}}K +a K\sqrt {\left( {\frac{{F^\prime  }}{{rF}}} \right)^2  - \frac{{G_{tr}^2 }}{{K^2 F^2 }}} \, .
\end{equation}
Here, $\phi ( {t,r} )$ is assumed to be a solution of the equations. Using the formula of $G_{tr}$;
\begin{equation}
\begin{split}
& G_{tr}  =  - \frac{{2\left( {r^3 H(t)\dot H(t) + \dot M(t)} \right)}}{{r\left( {r^3 H(t)^2  + 2M(t) - r} \right)}} =  - \frac{{2W\left( {t,r} \right)}}{{r^2 K}} \, , \\ 
\text{for}&\\
&\quad W\left( {t,r} \right) = r^3 H(t)\dot H(t) + \dot M(t) = \frac{r}{2}\frac{\partial}{{\partial t}}K \, , 
\end{split}
\end{equation}
we obtain
\[
8\pi \left. {L^{\left( \phi  \right)} } \right|_{\phi \left( {t,r} \right)} =  - 3H\left( t \right)^2  - \frac{{F^\prime  }}{{rF}}K + a K\sqrt {\left( {\frac{{F^\prime  }}{{rF}}} \right)^2  - \frac{{4W^2 }}{{r^4 K^4 F^2 }}}  \,  .
\]
This is assigning a value to the Lagrangian of $\phi$ in terms of metric functions.
\\

By that, the equation $G_{\theta \theta}= 8\pi g_{\theta \theta} L^{(\phi)}$ becomes
\begin{equation}\label{eq:9}
\begin{split}
  &- 3r^2 H(t)^2  - \frac{{r^2 K\,F''}}{F} - \left( {4rK - 9M(t) + 3r} \right)\frac{{F'}}{F} - \frac{1}{{K^3 F^2 }}\left( {rK\dot W - 4W^2 } \right) \\ 
  &+ \frac{1}{{K^2 F^3 }}\left( {r\dot F W} \right) =  - 3r^2 H\left( t \right)^2  - \frac{{rF^\prime  }}{F}K +a r^2 K\sqrt {\left( {\frac{{F^\prime  }}{{rF}}} \right)^2  - \frac{{4W^2 }}{{r^4 K^4 F^2 }}}  \, ,
\end{split}
\end{equation}
so
\begin{equation}\label{eq:10}
\begin{split}
&a\sqrt {\left( {\frac{{F^\prime  }}{{rF}}} \right)^2  - \frac{{4W^2 }}{{r^4 K^4 F^2 }}} \\
& =  - \frac{{\,F''}}{F} - 3\left( {1 + \frac{{ r - 3M }}{{rK}}} \right)\frac{{F'}}{{rF}} - \frac{1}{{r^2 K^4 F^2 }}\left( {rK\dot W - 4W^2 } \right)
 + \frac{{\dot F W}}{{rK^3 F^3 }}\,.
\end{split}
\end{equation}
This equation includes only the metric functions. As we will see (in the Appendix), it gives a differential equation for the function $F(t,r)$.
\\

Before searching for the solutions, we need to make the equations of $ \phi '$ and $\dot \phi$ (the equations (\ref{eq:6}) and (\ref{eq:14})) consistent, i.e, relate them to the same field $ \phi(t,r)$. We can do that by satisfying $\partial _t \phi ' = \partial _r \dot \phi $, but this gives a complicated equation that includes many roots. Instead of that, we can use the potential solution (\ref{eq:17}), to get (by taking the derivative with respect to $t$ and $r$) the equations
\begin{equation}\label{eq:a24}
\begin{split}
& \phi '\left( {t,r} \right) = \left( {8\pi \left. {\frac{d}{d\phi}V\left( \phi  \right)} \right|_{\phi \left( {t,r} \right)} } \right)^{ - 1} \left[ {\left( {\partial _r \frac{{F'(t,r)}}{{rF(t,r)}}} \right)K + \frac{{F'(t,r)}}{{rF(t,r)}}K'} \right]\, ,\\ 
& \dot \phi \left( {t,r} \right) = \left( {8\pi \left. {\frac{d}{d\phi} V\left( \phi  \right)} \right|_{\phi \left( {t,r} \right)} } \right)^{ - 1} \left[ {6H(t)\dot H(t) + \left( {\partial _t \frac{{F'(t,r)}}{{rF(t,r)}}} \right)K + \frac{{F'(t,r)}}{{rF(t,r)}}\dot K} \right]\, ,
\end{split}
\end{equation}
which relate $\phi '$ and $\dot \phi$ to the same field; that is the solution of
\[
\phi \left( {t,r} \right) = V^{ - 1} \left( {\left( {3H(t)^2  + \frac{{F'(t,r)}}{{rF(t,r)}}K} \right)/8\pi } \right)\, ,
\]
where $V^{ - 1}$ is the inverse function of the potential function $V$. 
\\

By using the equations (\ref{eq:a24}) in the equations (\ref{eq:6}) and (\ref{eq:14}), we end with vanishing the root term (see the Appendix);
\begin{equation}\label{eq:a29}
\sqrt {\left( {\frac{{F^\prime  }}{{rF}}} \right)^2  - \frac{{G_{tr}^2 }}{{K^2 F^2 }}}=0 \, ,
\end{equation}
which implies that
\begin{equation}\label{eq:a30}
W = \frac{1}{2}rK^2 F' \, .
\end{equation}
By that, we obtain more simple equations;
\begin{equation}\label{eq:a36}
\sqrt {8\pi } \phi '   = a\sqrt {\frac{{F'  }}{{rF}}} \,, \quad \text{and} \quad \sqrt {8\pi } \dot \phi   = b\sqrt {\frac{{FF^\prime  }}{r}} \, K \,,  
\end{equation}
for $a,b=\mp 1$. Actually, during the expansion of the universe, the amplitude of the field $ \phi$ is in decreasing with respect to time, therefore, we set $b= 1$ (where $K<0$).
\\

And by satisfying the consistence equation, $  \partial _t \phi ^\prime=\partial _r \dot \phi  $, for the equations (\ref{eq:a36}), we get the equation (see the Appendix)
\begin{equation}\label{eq:a31}
\frac{{ab}}{2}F\partial _t \left( {\frac{{F'  }}{{rF}}} \right) - \frac{{2FF^\prime  }}{{r^2 }}\left( {1 - \frac{{3M\left( t \right)}}{r}} \right) = \left( {\frac{1}{2}\partial _r \left( {\frac{{FF'  }}{r}} \right) + \frac{{2FF^\prime  }}{r^2}} \right)K\,.
\end{equation}
This equation gives the function $K(t,r)$ in terms of the function $F(t,r)$. 
\\

Getting the function $K(t,r)$ from the equation (\ref{eq:a31}) and using it with the equation (\ref{eq:a30}) in the equation (\ref{eq:10}), we end with (see the Appendix)
\begin{equation}\label{eq:a61}
\begin{split}
& \left( { \dot FF'  -  F\dot F'} \right)\left[ {F'^2  + \left( {\,FF'' + \frac{{3FF'}}{r}} \right)\left( {2ab + 1} \right)} \right] \\ 
& \quad \quad  \quad \quad + \frac{{2F^2 F'}}{{r^2 }}\left( {1 - \frac{{3M\left( t \right)}}{r}} \right)\left[ { - 3F'^2  + \,FF'' + \frac{{3FF'}}{r}} \right] = 0 \, . 
\end{split}
\end{equation}
This is a differential equation for the function $F(t,r)$. If we consider the mass of the black hole as almost stable; $M(t)=M_0+\delta M(t)$, for $\delta M(t)/M_0<<1$, we get an approximated equation;
\begin{equation}\label{eq:a32}
\begin{split}
& \left( {\dot FF'  -  F\dot F'} \right)\left[ {F'^2  + \left( {\,FF'' + \frac{{3FF'}}{r}} \right)\left( {2ab + 1} \right)} \right] \\ 
&\quad \quad  \quad \quad   + \frac{{2F^2 F'}}{{r^2 }}\left( {1 - \frac{{3M_0}}{r}} \right)\left[ { - 3F'^2  + \,FF'' + \frac{{3FF'}}{r}} \right] = 0 \, . 
\end{split}
\end{equation}
Actually, as we have mentioned before, that the function $F(t,r)$ has some physical role, that is, it contributes in spherical symmetric distributions of kinetic and potential energies over the space $(0, \infty) \times S^2$ at each moment of time. But during the expansion of the universe, the energy of the field $\phi$ decreases with respect to time, therefore $\left| F(t,r) \right |$ is also in decreasing with respect to time. That allows us to suggest some solutions for $F(t,r)$, as we will do in next sections. 
\\

\section{Solutions of slowly expanding universe }
In the previous section, we have obtained a differential equation for the function $F(t,r)$. By solving that equation, and by using that solution, we obtain the solutions of the field $\phi(t,r)$ and solutions of the metric functions, $M(t)$ and $H(t)$. Actually, the behaviour of the function $F(t,r)$ determines the behaviour of the global energy density; speed of its decreasing, and so speed of the expansion of the universe. In this work, we test two behaviours; one with powers of $1/t$ (slow expansion), and other with powers of $\exp(-t)$ (fast expansion). The both behaviours show that $\phi$ is concentrated (or created) inside the black hole (mostly in $r\approx 0$), as seen by both exterior and interior observers. The second behaviour (powers of $\exp(-t)$) can correspond to a non-vanishing ground state energy of $\phi$, and can give a constant rate $H(t)=H_0$ to the expansion of the universe (accelerating expansion), with rapidly decreasing in the universal energy density to a constant value (vacuum energy). We obtain also the solutions of $\phi(t,r)$ inside the black hole, with respect to the metric (\ref{eq:a69}), by using the swapping $t\to r$ and  $r\to t$, and find that $r=0$ (space point) is a singularity with respect to the exterior observers, while $t=0$ (time point) is a singularity with respect to the interior observers inside of the black hole. We see that we can avoid the divergence in solutions of $\phi(t,r)$ in $r=0$, but with respect to the interior observers. We find that the energy density of $\phi(t,r)$ inside the black hole (as measured by some interior observer) tends to take non-vanishing fixed values, and may increases (depending on the solutions of $F(t,r)$).
\\
\\

In this section, we study the behaviour of solutions of $1/t$-power series. Let us expand $F(t,r)$ as 
\begin{equation}\label{eq:a33}
\begin{split}
 F\left( {t,r} \right) = f_0 \left( r \right) + \frac{{f_1 \left( r \right)}}{t} + \frac{{f_2 \left( r \right)}}{{t^2 }} + ...  = \sum\limits_{n \ge 0} {f_n \left( r \right)t^{ - n} }  \, , 
\end{split}
\end{equation}
for some differentiable functions $\{ f_n ( r )\}$ which can be obtained by the identification with respect to powers of $1/t$. Actually, at late times of the expansion of the universe, we have $t>>1$, therefore the series (\ref{eq:a33}) indeed converges, and $F(t,r)$ limits to a static solution, $f_0 ( r ) $, with respect to the exterior observers. We note that we can shift $t$ to $t+t_0$ (for $t_0=\text{constant}$) without changing the functions $\{ f_n ( r )\}$.
\\

Using the series (\ref{eq:a33}) in the equation (\ref{eq:a32}) and identify with respect to the powers of $1/t$, we obtain
\begin{equation}\label{eq:a35}
 - 3f_0 ' \left( r \right)^2  + \,f_0 \left( r \right)f_0 ''\left( r \right) + \frac{3}{r}f_0 \left( r \right)f_0 '\left( r \right) = 0\, ,
\end{equation}
and
\begin{equation}\label{eq:a34}
\begin{split}
& \sum\limits_{n,k \ge 0} {\left( {k - 2n - 1} \right)f_n \left( r \right)f_{k - n - 1} '\left( r \right)a_{m - k} \left( r \right)}  \\ 
  &+ \frac{2}{{r^2 }}\left( {1 - \frac{{3M_0 }}{r}} \right)\sum\limits_{n,k,l \ge 0} {f_n \left( r \right)f_{k - n} \left( r \right)f_{l - k} '\left( r \right)b_{m - l} \left( r \right)}\,  ,\quad \text{for} \quad m\ge 1 \, . 
\end{split}
\end{equation}
Where
\begin{equation}\label{eq:a44}
\begin{split}
& a_k \left( r \right) = \sum\limits_{n \ge 0} {f_n '\left( r \right)f'_{k - n} \left( r \right)}  + \left( {2ab + 1} \right)\left( {\,f_n \left( r \right)f_{k - n} ''\left( r \right) + \frac{3}{r}f_n \left( r \right)f_{k - n} '\left( r \right)} \right) \, ,\\ 
\text{and}&\\
& b_k \left( r \right) = \sum\limits_{n \ge 0} { - 3f_n '\left( r \right)f'_{k - n} \left( r \right)}  + \,f_n \left( r \right)f_{k - n} ''\left( r \right) + \frac{3}{r}f_n \left( r \right)f_{k - n} '\left( r \right) \, . 
\end{split}
\end{equation}
The equation (\ref{eq:a35}) has the general solution
\begin{equation}\label{eq:a42}
f_0 (r) = \frac{{c_1 r}}{{\sqrt {r^2  + c_2 } }} \, ,
\end{equation}
for some arbitrary constants $c_1$ and $c_2>0$. We note that the function $f_0 (r)$ satisfies the condition $f_0 (r) f_0' (r)>0$, therefore the function $F (t,r)$ also satisfies this condition, but at late times of the expansion of the universe, where the function $F(t,r)$ approximates to the function $f_0 (r)$.
\\

For $m=1$, the equation (\ref{eq:a34}) gives
\begin{equation}
\begin{split}
f_0 \left( r \right)f_1 ''\left( r \right) + \left( {\frac{3}{r}f_0 \left( r \right) - 6f_0 '\left( r \right)} \right)f_1 '\left( r \right) + \left( {\frac{3}{r}f_0 '\left( r \right) + \,f_0 ''\left( r \right)} \right)f_1 \left( r \right) = 0 \, ,
\end{split}
\end{equation}
which determines the function $f_1 ( r )$;
\[
f_1 \left( r \right) = \frac{{c_3 r + c_4 \,r^3 }}{{\left( {r^2  + c_2 } \right)^{\frac{3}{2}} }} \, ,
\]
for some arbitrary constants $c_3$ and $c_4$. And for $m=2$, it gives
\[
\left( {6ab + 4} \right)f_0 '\left( {f_0 f_1 ' - f_0 'f_1 } \right) + \frac{2}{{r^2 }}\left( {1 - \frac{{3M_0 }}{r}} \right)f_0^2 b_2  = 0\, ,
\]
where
\begin{equation}
\begin{split}
 b_2&  = f_0 f_2 '' + \left( { - 6f_0 ' + \frac{3}{r}f_0 } \right)f_2 ' + \left( {f_0 '' + \frac{3}{r}f'_0 } \right)f_2  \\ 
  &- 3f_1 '^2  + f_1 f_1 '' + \frac{3}{r}f_1 f_1 ' \, . 
\end{split}
\end{equation}
The last two equations determine the function $f_2(r)$ in terms of the functions $f_0(r)$ and $f_1(r)$. And like that, we obtain the remaining functions $f_3(r), f_4(r), ...$ .
\\

We can get the solutions of the field $\phi ( {t,r} )$ by using only one of the equations (\ref{eq:a36}), this is right because we have already imposed the consistence condition, $\partial_t \phi' =\partial_r\dot \phi$. By integration equation of $\phi'$, we obtain
\begin{equation}\label{eq:a63}
\sqrt {8\pi } \phi \left( {t,r} \right) =  - a\ln \left( {\frac{{2\left( {\sqrt 6 \,\sqrt {r^2  + 6}  + 6} \right)}}{r}} \right) + \frac{{\sqrt 6 \,\left( {6c_4  - c_3 } \right)}}{{4\left( {r^2  + 6} \right)^2 }}\frac{1}{t}+\mathcal{O}(1/t^2)+\phi_0 ( {t} )+c \, ,
\end{equation}
where the constant $c=+a\ln(2\sqrt 6)+\cdots $ is needed for obtaining $\phi \left( {t,r} \right) \to 0$ at $r>>1$, and the function $ \phi_0(t)$ is an arbitrary and can be obtained by satisfying $\dot \phi(t,r)=0$ on $K=0$ surfaces. We have set $c_2=6$ (this value will be obtained in next sections). 
\\

Actually, this solution of $\phi ( {t,r} )$ approaches to a time independent solution;
\[
\sqrt {8\pi } \phi \left( {t,r} \right) \to  - \,a\ln \left( {\frac{{2\left( {\sqrt 6 \,\sqrt {r^2  + 6}  + 6} \right)}}{r}} \right) \, ,
\]
at late times of the expansion of the universe, $t>>1$. The behaviour of this solution (for $r>R_{BH}$) is represented in figure (\ref{fig:1}).
\begin{figure}[H]\label{fig:1}
  \includegraphics[width=0.6\textwidth]{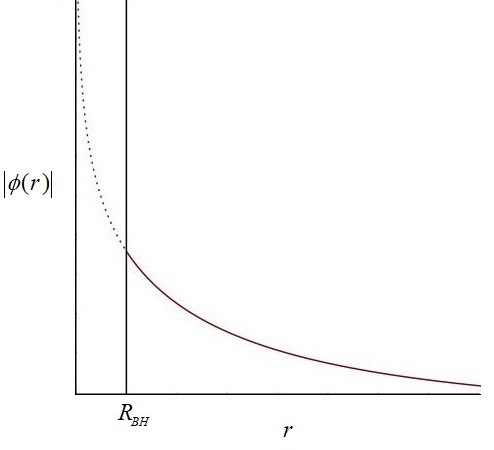}
  \caption{The inferred behaviour of $\phi(r)$ from exterior point of view.}
\label{fig:1}
\end{figure}

It is clear in the figure (\ref{fig:1}), with respect to the exterior observers ($r>R_{BH}$), that the amplitude of the field $\phi$ increases rapidly by getting closer to the singularity of the space-time. This behaviour can be seen as a concentration (or creation) of $\phi$ inside that black hole (or nearby it) due to the gravity attractive force of that black hole. In other words, the incoming waves of $\phi$ towards the black hole can not escape back away from it (as seen by exterior observers). 
\\

But inside the black hole ($r<R_{BH}$), as discussed above equation (\ref{eq:a70}), the interior observers need to swap the coordinates; $t$ to $r$ and  $r$ to $t$, therefore, we swap the solution $\phi(t,r)$ to $\phi(r,t)$, so the solution (\ref{eq:a63}) changes to
\begin{equation}\label{eq:a64}
\begin{split}
\left. {\sqrt {8\pi } \phi \left( {t,r} \right)} \right|_{\text{interior} }  = & -a\ln \left( {\frac{{2\left( {\sqrt 6 \,\sqrt {t^2  + 6}  + 6} \right)}}{t}} \right) + \frac{{\sqrt 6 \,\left( {6c_4  - c_3 } \right)}}{{4\left( {t^2  + 6} \right)^2 }}\frac{1}{r} \\
&    +\mathcal{O}(1/r^2)+\cdots \,,
\end{split}
\end{equation}
which is the solution with respect to the metric (\ref{eq:a69}).
\\

We note that the approximated differential equation of the function $F$, equation (\ref{eq:a32}), is invariant under an arbitrary shifting in the time; $t\to t+t_0$ (for $t_0=\text{constant}$) \footnote{As mentioned under the equation (\ref{eq:a33}), the shifting $t\to t+t_0$ is possible and can be regarded as a transformation of the variable $t$, because the variable $t$ does not appear explicitly in the approximated differential equation of the function $F$.}, that implies a shifting in $r$ in the solution (\ref{eq:a64}), to obtain
\begin{equation}\label{eq:a65}
\begin{split}
\left. {\sqrt {8\pi } \phi \left( {t,r} \right)} \right|_{\text{interior} }  =&  -a\ln \left( {\frac{{2\left( {\sqrt 6 \,\sqrt {t^2  + 6}  + 6} \right)}}{t}} \right) + \frac{{\sqrt 6 \,\left( {6c_4  - c_3 } \right)}}{{4\left( {t^2  + 6} \right)^2 }}\frac{1}{r+r_0}  \\
  & +\mathcal{O}(1/(r+r_0)^2)+\cdots \,.
\end{split}
\end{equation}
This shifting is needed to keep the solutions convergent (by choosing enough large constant $r_0$) inside the black hole, where the intervals of time and space are bounded. The shifting $r\to r+r_0$ also shows that $r=0$ is not a singularity with respect to the observes inside the black hole, while $t=0$ is the singularity, but $t=0$ has been passed because of $dt>0$ in the swapped coordinates (with respect to the interior metric (\ref{eq:a69})). 
\\

According to the solution (\ref{eq:a65}), the observers inside the black hole note that the field $\phi$ is concentrated around the point $r=0$, but it does not divergent. And because the intervals of the swapped $t$ and $r$ are bounded (to satisfy $g_{rr} > 0$), the absolute value of the solution (\ref{eq:a65}) is bounded from below and does not vanish, i.e, it tends to take a non-vanishing constant value when the swapped $t$ and $r$ take their final values. That means that the field $\phi$ inside the black hole (as seen by interior observers) is preserved (stable) under influences of the expansion of the universe. To see that, let us consider an observer that located inside the black, in some $r=\text{constant}$ ($\dot r=0$). Because the time inside the black hole is limited, approximately $t\in (0, 2M(t)) \to (0, 2M_0)$, the velocity in $t$-coordinate should get infinitesimal values by getting closer to $t_{max}=2M_0$, till vanishes. Therefore, we can write that velocity as 
\[
u = \left( { - 1 + \frac{{2M(r)}}{t} + t^2 H(r)^2 ,0,0,0} \right);\quad u^2  = g_{\mu \nu } u^\mu  u^\nu   = 1 - \frac{{2M(r)}}{t} - t^2 H(r)^2 \le 0\, ,
\]
where we used the interior metric (\ref{eq:a69});
\begin{equation}
\begin{split}
\left. {ds^2 } \right|_{\text{interior BH} }   = \frac{1}{{1 - \frac{{2M(r)}}{t} - t^2 H(r)^2 }}dt^2  &+ \left( { - 1 + \frac{{2M(r)}}{t} + t^2 H(r)^2 } \right)F(r,t)^2 dr^2  \\
&+ t^2 \left( {d\theta ^2  + \sin ^2 (\theta )d\Phi ^2 } \right)\,.
\end{split}
\end{equation}
By that, the velocity $u^t \ge 0$ vanishes on the black hole horizon, so that observer needs infinite proper time to reach that horizon (boundary), in other words, the time $t_{max}=2M_0$ inside the black hole corresponds to $t=\infty$ outside of it. Let us calculate the energy density of the field $\phi$, as measured by this observer. From the formula of $T_{\mu \nu}$ (the equation (\ref{eq:a67})), we obtain the energy density;
\begin{equation}\label{eq:a68}
\begin{split}
 \rho ^{\left( \phi  \right)} _{\text{interior BH}}  = T^{\left( \phi  \right)} _{\mu \nu } u^\mu  u^\nu   = &\frac{1}{2}\left( { - 1 + \frac{{2M(r)}}{t} + t^2 H(r)^2 } \right)^2 \dot \phi ^2_{in}  + \frac{1}{2}\frac{{\phi '^2_{in} }}{{F(r,t)^2 }} \\ 
  &+ \left( { - 1 + \frac{{2M(r)}}{t} + t^2 H(r)^2 } \right)V\left( \phi_{in}  \right) \, . 
\end{split}
\end{equation}
Actually, the same expression for $ \rho ^{( \phi  )} _{\text{interior}} $ can be obtained by using the definition $ \rho  \equiv T^{00} $. Regarding the fact that the field $\phi|_{\text{interior} }$ and its derivatives are convergent ($t>0$), this energy density also converges and tends to take a non-vanishing fixed values at $t=t_{max}\approx 2M_0$. Actually, $ \rho ^{( \phi  )} _{\text{interior}} $ vanishes on the black hole horizon surface, where $\phi '^2|_{\text{interior} }$ also vanishes on that surface. 
\\

Using the equations (\ref{eq:25}) and (\ref{eq:17}), with swapping $t\to r$ and $r\to t$, the equation (\ref{eq:a68}) becomes
\begin{equation}
\begin{split}
 8\pi \rho _{{\text{interior BH}}}^{\left( \phi  \right)}  &= \frac{ 8\pi}{{F(r,t)^2 }}\left( {\frac{1}{2}K(r,t)^2 F(r,t)^2 {\dot \phi} _{in}^2  + \frac{1}{2}{\phi '}_{in}^2 } \right) +  8\pi K(r,t)V\left( {\phi _{in} } \right)\, \\ 
 & = \frac{1}{{F(r,t)^2 }}\frac{{F(r,t)\dot F(r,t)}}{t}K(r,t)^2  + K(r,t)\left( {3H(r)^2  + \frac{{\dot F(r,t)}}{{tF(r,t)}}K(r,t)} \right)\, \\ 
 & = \frac{{2\dot F(r,t)}}{{tF(r,t)}}K(r,t)^2  + 3H(r)^2 K(r,t) \, .
\end{split}
\end{equation}
We note that the behaviour of the function $F_{\text{interior}} (t,r)=F(r,t)$ can control the behaviour of $ \rho ^{( \phi  )} _{\text{interior}} $; If $ \dot F/F$, inside the black hole, increases enough when $t\to 2M_0$ (late times of expansion), then the energy density $ \rho ^{( \phi  )} _{\text{interior}} $ indeed increases, which means that there is an energy accretion from the exterior of the black hole into its interior. The conditions on the function $F$, in $r> R_{BH}$, are that $FF'>0$ and $|F|$ needs to decrease with respect to the time ($F \dot F<0$). Inside the black hole, $r< R_{BH}$, the condition $FF'>0$ swaps to $F \dot F>0$, which gives the possibility of getting that $ \dot F/tF$ increases with respect to the time for some solutions of the function $F$, so giving the possibility of getting that $\dot \phi ^2_{in}$ and $\phi '^2_{in}$ increase with respect to the time. We have the approximation for $K(r,t)$ inside the black hole:
\[
K(r,t) = t^2 H(r)^2  + \frac{{2M(r)}}{t} - 1 \approx \frac{{2M(r)}}{{2M_0 }} - 1 \approx \frac{{2\left( {M(R_{BH} ) + M'(R_{BH} )\Delta r} \right)}}{{2M_0 }} - 1 \, ,
\]
where we ignored $t^2 H(r)^2 <<1$, and for $R_{BH}\to 2M_0$ at late times of cosmological expansion (with taking in consideration the shifting $r\to r+r_0$ as did in equation (\ref{eq:a65})), we obtain
\[
8\pi \rho _{{\text{interior BH}}}^{\left( \phi  \right)}  \approx \frac{{2\dot F(r,t)}}{{tF(r,t)}}\left( {\frac{{M'(R_{BH} )}}{{M_0 }}\Delta r} \right)^2  + 3H(r)^2 \frac{{M'(R_{BH} )}}{{M_0 }}\Delta r \, ,
\]
where $\Delta r= R_{BH}-r >0$. Using the volume $4\pi t^3 /3 $ (as can be defined by the metric (\ref{eq:a69})), we obtain an approximated energy of $\phi$ inside of the black hole, $r< R_{BH}$;
\[
8\pi E _{{\text{interior BH}}}^{\left( \phi  \right)}  \approx \frac{{8\pi \dot F(r,t)}}{{3F(r,t)}}\left( {\frac{{M'(R_{BH} )}}{{M_0 }}\Delta r} \right)^2 t^2 + {4\pi} H(r)^2 \frac{{M'(R_{BH} )}}{{M_0 }}\Delta r t^3 \, ,
\]
this energy increases with respect to the time $t\to 2M_0$, this implies an assertion of energy from the exterior of the black hole into its interior, thus, there is a possibility of growth of the black holes by growing the real scalar field inside of it.
\\

We note that both observers inside and outside the black hole note that the field $\phi$ is concentrated around the point $r=0$ (centre of BH mass), therefore, that concentration is a global phenomenon. 
\\
\\

We can get the formula of the cosmological horizon radius, $r=R_C(t)$, as a function of time, by using the equations
\[
\sqrt {8\pi } \phi ^\prime   = a\sqrt {\frac{{F^\prime  }}{{rF}}} \,,\quad \text{and}\quad \sqrt {8\pi } \dot \phi  = b\sqrt {\frac{{FF^\prime  }}{r}} \,K\, ,
\]
and by applying the consistence condition; $\partial_r \dot\phi=\partial_t \phi ^\prime  $, we obtain the equation
\[
b\left( {\partial _r \sqrt {\frac{{FF^\prime  }}{r}} } \right)\,K + b\sqrt {\frac{{FF^\prime  }}{r}} \,K' = a\partial _t \sqrt {\frac{{F^\prime  }}{{rF}}} \, .
\]
On cosmological horizon surface, $r=R_C(t)$ (where $K=0$), this equation becomes
\begin{equation}\label{eq:a37}
b\left. {\sqrt {\frac{{FF'}}{r}} \,K'} \right|_{r = R_C \left( t \right)}  = a\left. {\partial _t \sqrt {\frac{{F'}}{{rF}}} } \right|_{r = R_C \left( t \right)} \, .
\end{equation}
This equation can simply give the solution of $r=R_C(t)$. At late times of expansion of the universe, the function $F(t,r)$ approximates to the static function $f_0 (r)$, according to the power series (\ref{eq:a33}). By that, the formula (\ref{eq:a37}) gives
\[
b\left. {\frac{{c_1 \sqrt {c_2 } }}{{r^2  + c_2 }}\,K'} \right|_{r = R_C \left( t \right)}  = a\left. {\frac{{\left( {c_2 c_4  - c_3 } \right)}}{{c_1 \sqrt {c_2 } }}\frac{{ - r}}{{\left( {r^2  + c_2 } \right)^{\frac{3}{2}} }}} \right|_{r = R_C \left( t \right)} t^{ - 2}  + \mathcal{O}(t^{ - 3} )\, ,  for\, \,  t>>1\, ,
\]
therefore
\begin{equation}\label{eq:a38}
\begin{split}
 \alpha\left. {\,K'} \right|_{r = R_C \left( t \right)}  =  \frac{{ R_C }}{{\left( {R_C^2  + c_2 } \right)^{\frac{1}{2}} }}t^{ - 2}  + \mathcal{O}(t^{ - 3} ) \, ,\quad  \text{for}\quad \alpha  =-ab \frac{{{\left( c_1 \sqrt {c_2 } \right)^2}}}{{\left( {c_2 c_4  - c_3 } \right)  }} \, . 
\end{split}
\end{equation}
Actually, the function $K'=2r H^2 -2 M/r^2$ includes two functions, $H(t)$ and $M(t)$, but at late times of the expansion of the universe, we have $M/R_C<<1$. Therefore, we can use the approximation
\[
\left. {\,K'} \right|_{r = R_C \left( t \right)}  = 2R_C H^2 \, ,
\]
and by using the approximation $H\approx 1/R_C$, the equation (\ref{eq:a38}) implies
\[
\alpha\frac{2}{{R_C }} =  \frac{{  R_C }}{{\left( {R_C^2  + c_2 } \right)^{\frac{1}{2}} }}t^{ - 2}  +  \mathcal{O}(t^{ - 3} ) \, .
\]
Taking in consideration that during the expansion of the universe, the universal energy density decreases, therefore $H(t)$ also decreases, so $R_C(t)$ increases. Therefore, for any value of $c_2$, it will be that $R_C^2  >> c_2$ at late times of cosmological expansion, so we can ignore $c_2$ in the previous formula and get
\[
R_C \left( t \right) =   2\alpha t^2 \, .
\]
As expected, $R_C(t)>0$ increases with respect to time during the expansion of the universe. By that, the expansion rate $H\approx 1/R_C$ becomes
\begin{equation}\label{eq:a39}
H\left( t \right) = \,\frac{{\left( {2\alpha } \right)^{ - 1} }}{{t^2 }} = H\left( {t_0 } \right)\,\left( {\frac{{t_0 }}{t}} \right)^2 \, , \quad \text{for}\quad t>>t_0\, .
\end{equation}
Therefore, the universal energy density, $3H^2$, decreases to infinitesimal values, at late times of expansion of the universe. 
\\

Next, we get the mass function of the black hole. We note that there is a kind of energy conservation between the black hole and the field $\phi$, on horizon surfaces; The equations (\ref{eq:6}) and (\ref{eq:14}) with $G_{tr}  =8\pi \dot \phi \phi'$ show that $W( {t,r} ) = r^3 H(t)\dot H(t) + \dot M(t)\approx0$ nearby horizon surfaces, $K=0$ ($r = R_H)$. Therefore, nearby black hole horizon surface $r \approx R_{BH}$, we obtain
\begin{equation}\label{eq:27}
 R_{BH}^3 H(t)\dot H(t) + \dot M(t)\approx0, \quad \text{for} \quad K\approx0 \, .
\end{equation}
If we think that this term indeed vanishes on $K=0$ surfaces, it implies that
\[
 \frac{{4\pi }}{3}R_{BH}^3 \dot \rho _c  + 8\pi \dot M(t) = 0 \, ,  
\]
where $ \rho _c  = 3H^2 $ is the universal critical energy density (gravitational energy), while $4\pi R_{BH}^3/3$ represents the volume inside of the black hole horizon surface ($r=R_{BH}$) as as seen by the exterior observers, this volume is proposed in \cite{Maulik} \footnote{In fact, the notation of the ''volume'' for the black holes is still unclear, for example, there are other definitions of the volume of the black hole, as proposed in \cite{Marios}.}. Therefore, the quantity $ \dot \rho _c 4\pi R_{BH}^3/3 $ represents the changes of the gravitational energy inside the volume $4\pi R_{BH}^3/3$, when $ R_{BH}=\text{constant}$. That means that there is an exchanging energy between the black hole and the gravitational energy. But during the expansion, it is always be $\dot H(t)<0$, which implies that $\dot M(t)>0$ is always satisfied. This means that the expansion of the universe influences increasing in the mass of black hole, but this increasing is slightly since $\dot H(t)$ becomes small at late times of the expansion, and so the mass of the black hole is almost stable in view of expansion of the universe. We have also $V(\phi)=3H^2$ on horizon surfaces, this implies
\[
 \frac{{4\pi }}{3}R_{BH}^3 \dot V(\phi)  + 8\pi \dot M(t) = 0 \, .
\]
This clearly shows that there is an exchanging energy between the black hole and the scalar field.
\\

We can obtain the mass function of the black hole, $M(t)$, by using the approximation $R_{BH}(t) \approx 2M(t)$ in equation (\ref{eq:27}), to get
\begin{equation}
\left( {2M(t)} \right)^3 H(t)\dot H(t) + \dot M(t) = 0 \, ,
\end{equation}
which has the general solution
\begin{equation}\label{eq:a45}
M(t)^2  = \frac{{M_0^2 }}{{8M_0^2 H\left( {t } \right)^2   + 1}} \,  ,
\end{equation}
for some constant mass $M_0$. It is clear that the mass of the black hole increases with respect to the time during the expansion of the universe, where the universal critical energy density $3H( {t } )^2$ decreases. 
\\

By using solution of $H(t)$, equation (\ref{eq:a39}), we obtain
\[
M(t)^2  = \frac{{M_0^2 }}{{8M_0^2 H\left( {t_0 } \right)^2 \,\left( {\frac{{t_0 }}{t}} \right)^4  + 1}} \approx M_0^2 \left( {1 - 8M_0^2 H\left( {t_0 } \right)^2 \,\left( {\frac{{t_0 }}{t}} \right)^4 } \right) \, .
\]
This shows that the mass of the black hole increases with respect to time during the expansion of the universe, but slightly. And at late times of that expansion, the mass of the black hole tends to take a constant value although the global energy density tends to take infinitesimal values due to the expansion. Therefore, the black hole is stable in view of the expansion of the universe.
\\

As we have mentioned in the introduction, astronomically, we can regard the black holes as cosmological objects without horizons and without singularities, so we can use RW cosmology and write the critical energy density $3H( {t } )^2$ as a sum of the universal energy components, to get
\begin{equation}
\begin{split}
H(a)^2 {\rm{  }} = H_0^2 \left( {\Omega _{r0} a^{ - 4}  + \Omega _{m0} a^{ - 3}  + \Omega _{de0} a^{ - 3\left( {1 + \omega _\phi  } \right)} } \right)\, = H_0^2 \sum\limits_i {\Omega _{i0} a^{ - k_i } } \,,
\end{split}
\end{equation}
where $\omega _\phi$ is the equation of state of the field $\phi$, and $a(t)$ is RW scale factor. If the universe is dominated by a component '$i$', and for early times of the cosmological expansion; $M_0 H(t) >>1$, the equation (\ref{eq:a45}) approximates to
\begin{equation}\label{eq:a55}
M(a) \approx \frac{1}{{\sqrt {8\Omega _{i0}} H_0  }} \left( {\frac{a}{{a_0 }}} \right)^{\frac{k_i}{2}}  = M(a_i )\left( {\frac{a}{{a_i }}} \right)^{\frac{k_i}{2} } \,.
\end{equation}
While at late times of the expansion, it becomes $M_0 H(t) <<1$, so we get
\[
 M(t) \approx M_0 \left( {1 - 4M_0^2 H\left( t \right)^2 \,} \right)\, ,
\]
which implies that
\[
M(a) \approx M_0 - \alpha _{r0} \left( {\frac{a}{{a_0 }}} \right)^{ - 4}  - \alpha _{m0} \left( {\frac{a}{{a_0 }}} \right)^{ - 3}  - \alpha _{de0} \left( {\frac{a}{{a_0 }}} \right)^{ - 3\left( {1 + \omega _\phi  } \right)}  \, ,
\]
for some constants $\alpha _{i0}$. In both cases, the mass $M(a)$ increases with respect to $a$. And finally it approaches some fixed value $M_0$ although the universal energy density decreases to infinitesimal values.

\section{Solutions of fast expanding universe; Accelerated expansion}
In this section, we show that there are solutions that allow us to obtain a fixed rate for the cosmological expansion; $H(t)=H_0=\text{constant}$, but at late time of that expansion. That fixed rate of cosmological expansion associates with a rapidly decreasing in the universal energy density to reach a non-vanishing constant value which is the ground state energy density of the field $\phi$. That solution associates with a constant black hole mass, $M(t)=M_0=\text{constant}$.
\\

Actually, the differential equation of the function $F(t,r)$, the equation (\ref{eq:a32}), has a special solution, that is $F(t,r)=\text{constant}$, this implies that $\dot F(t,r)=0$ and $F'(t,r)=0$. If we use these equations in equation of $W(t,r)$, the equation (\ref{eq:a30}), we obtain
\[
r^3 H(t)\dot H(t) + \dot M(t) = 0\,,  \quad \text{for all}\, \,  r\ne 0 \,.
\]
This implies that $\dot H(t)=0$ and $\dot M(t) = 0$. And if we use the conditions, $\dot F(t,r)=0$ and $F'(t,r)=0$, in the equations 
\[
\sqrt {8\pi } \phi ^\prime   = a\sqrt {\frac{{F^\prime  }}{{rF}}} \,,\quad \text{and}\quad \sqrt {8\pi } \dot \phi  = b\sqrt {\frac{{FF^\prime  }}{r}} \,K\,,
\]
it implies that $\phi(t,r)=\phi_0=\text{constant}$. While, if we use them in the potential equation (\ref{eq:17});
\[
8\pi V\left( \phi  \right)   =3H(t)^2+  \frac{{ {F'(r)}  }}{{rF(r)}} K ; \quad K\le 0\, ,
\]
it implies that $V( \phi  ) |_{\phi=\phi_0}  =3H_0=\text{constant}$. This special solution corresponds to a vacuum energy that is given by $V( \phi_0  )  =\text{constant}\ne 0$ for some non-vanishing ground state expectation value of the field $\phi$. This behaviour can be obtained by using solutions with power series in $\exp(-t)$.
\\

As we mentioned before, that the function $F(t,r)$ has some physical role, that is, it contributes in spherical symmetric distributions of kinetic and potential energies over the space $(0, \infty) \times S^2$ at moment of time. But during the expansion, the energy of the field $\phi$ decreases with respect to time, therefore the function $\left | F(t,r) \right |$ is also in decreasing with respect to time. 
\\

Actually, a rapidly decreasing behaviour can be obtained by using the exponential function $\exp(-t)$. Let us write the function $F(t,r)$ as a power series in $\exp(-t)$, like
\begin{equation}\label{eq:a43}
 F\left( {t,r} \right) = c + f_1 \left( r \right)e^{ - t}  + f_2 \left( r \right)e^{ - 2t}  + ... = \sum\limits_{n \ge 0} {f_n \left( r \right)e^{ - nt} } \,,  
\end{equation}
 for $f_0 ( r ) = c=\text{constant}\ne0$. By that the function $F(t,r)$ decreases to reach a constant value, $c$, but rapidly. Using the equation (\ref{eq:a43}) in the differential equation of $F(t,r)$, the equation (\ref{eq:a32}), and identify with respect to power of $\exp(-t)$, we obtain
\begin{equation}
\begin{split}
& \sum\limits_{n,p\ge 0} {\left( {2n - p} \right)f_{p - n} \left( r \right)f'_n \left( r \right)a_{m - p} \left( r \right)}  \\ 
&  + \frac{2}{{r^2 }}\left( {1 - \frac{{3M_0 }}{r}} \right)\sum\limits_{n,p,l \ge 0} {f_n \left( r \right)f_{p - n} \left( r \right)f'_{l - p} \left( r \right)b_{m - l} \left( r \right)} \,,\quad \text{for}\quad m \ge 0\,,
\end{split}
\end{equation}
where the functions $a_n(r)$ and $b_n(r)$ are same as defined in (\ref{eq:a44}).
\\

Actually, for $m=0,1,2$, we obtain the same equation;
\[
f''_1 \left( r \right) + \frac{3}{r}f'_1 \left( r \right) = 0 \, ,
\]
which has the general solution
\[
f_1(r)={c_1}-\frac{c_2}{r^{2}} \, ,
\]
for some constants $c_1$ and $c_2>0$. While for $m=3$, we obtain equation of $f_2 ( r )$;
\[
f_2 \left( r \right) = \frac{{4c_2^2 }}{c}\int {\left( {\frac{1}{{r^3 }}\int {\frac{{r^3  + 18cM_0  - 6rc}}{{r^3\left ( - \left( {2ab + 1} \right)\,r^3  + 6cM_0  - 2rc\right)}}} dr + c_3 } \right)} dr + c_4  \, ,
\]
for some constants $c_3$ and $c_4$. And like that, we can obtain the remaining functions $f_3 ( r )$, $f_4 ( r )$,... . 
\\

By using these solutions in the equation (\ref{eq:a37}), and following same steps, we obtain 
\[
R_C \left( t \right) =  { - 4abc  - 4ab f'_1{^ {- 1}} \left( {2f_2 f_1 ' - cf'_2 }\right) e^{ - t}  + \mathcal{O}\left( {e^{ - 2t} } \right)}  \, .
\]
Actually, the obtained functions $f_1 ( r )$ and $f_2 ( r )$ decrease with respect to $r$, so they dot not effect efficiently on the this solution of $R_C$. Thus, at late time of the expansion, $t>>1$, the cosmological horizon radius $R_C(t)$ tends to take a constant value;
\[
R_C \left( t \right) =  - 4abc=\text{constant} \, .
\]
Because the constant $c$ is positive, $ab$ is negative, so $ab=-1$, and $R_C(t)=4c$. \\

By that, the expansion rate, $H(t)$, also tends to take a constant value; $H(t)=H_0=1/4c$, so obtaining accelerated expansion of the universe. A same thing we obtain for the solutions of the mass of the black hole (equation (\ref{eq:a45})), the field $\phi$ and its potential $V(\phi)$, they tend to take a constant non-vanishing values, but at late time of the expansion, $t>>1$. And because the solutions $f_n (r)$ are concentrated around $r=0$, so the solutions of $\phi$ are also concentrated around $r=0$.
\\

Inside the black hole, as discussed under the figure (\ref{fig:1}), we swap $t$ to $r$ and $r$ to $t$, in the solution (\ref{eq:a43}), and by using that solution in equation of $\phi'$ (which swaps to $\dot \phi$), we obtain the solutions of $\phi$ inside the black hole. We note that in the both cases, slow and fast expansion, the solutions of the field $\phi$ are concentrated (or created) around $r=0$ inside the black hole (mostly in $r\approx 0$).

\section{Solutions of the potential $V( \phi  )$ in terms of the function $F(t,r)$}
In the previous sections, we have obtained the solutions of the metric functions $F(t,r)$, $H(t)$ and $M(t)$, at late times of expansion of the universe, $t>>1$. But in the case of slow expansion of the universe, the function $F(t,r)$ limits to a static function, $f_0(r)$, as seen by the exterior observers (located at $r>R_{BH}$). In this section, we obtain a simple solution for $\phi(t, r)$ nearby horizon surfaces, and use it in getting a relation between the function $f_0(r)$ and the potential $V( \phi  )$, so determining the potential in terms of the function $f_0(r)$. Vice versa, we can obtain the function $f_0(r)$ by using a given potential $V( \phi  )$, and after obtaining $f_0(r)$, we write $F(t,r)=f_0(r) + F_1(t,r)$ and use it in the differential equation of $F(t,r)$ to obtain a differential equation for $F_1(t,r)$ which has to satisfy $F_1(t,r) \to 0$ at $t>>1$. 
\\

According to the potential equation (\ref{eq:17});
\[
8\pi V\left( \phi  \right)   =3H(t)^2+  \frac{{ {F'(r)}  }}{{rF(r)}} K ; \quad K\le 0\, ,
\]
and on horizon surfaces, $K=0$, this potential takes the value
\begin{equation}\label{eq:a41}
8\pi \left. {V\left( \phi  \right)} \right|_{r = R_H }  = 3H(t)^2 \, .
\end{equation}
But since the right side of this equation depends only on time, this allows us to suggest a solution for the field $\phi$ nearby horizon surfaces by writing 
\[
\phi \left( {t,r} \right) = \phi _0 \left( t \right) + \phi _1 \left( {t,r} \right),\quad \text{where} \quad \left. {\phi _1 \left( {t,r} \right)} \right|_{r = R_H }  = 0\, .
\]
And since it becomes $ \phi _1 <<  \phi _0$ by getting closer to Horizons surfaces ($r=R_H$), we can expand the potential like
\[
V\left( \phi  \right) = V\left( {\phi _0  + \phi _1 } \right) = V\left( {\phi _0 } \right) + V'\left( {\phi _0 } \right)\phi _1  + ... \, ,
\]
which as required satisfies
\begin{equation}
8\pi \left. {V\left( \phi  \right)} \right|_{r = R_H }  = 8\pi \left. {V\left( {\phi _0 } \right)} \right|_{r = R_H }  = 3H(t)^2 \, .
\end{equation}
The solution $\phi _0(t)$ is a global, since it does not depend on the space coordinates $( r, \theta, \Phi) $, it gives a similarity with spatially isotropic homogenous space-time, the RW space-time, thus it can describe the global dark energy which has same density over the universe. 
\\

Actually, we can get the solution of $\phi ( {t,r} )=\phi _0(t)+ \phi _1(t,r)$ by using it in the formula of $\phi '  $, the equation (\ref{eq:a36}), we obtain
\begin{equation}\label{eq:a62}
\sqrt {8\pi } \phi' _1 \left( t,r \right)  \approx a\sqrt {\frac{{f'_0 \left( r \right)}}{{rf_0 \left( r \right)}}} \,,\quad \text{at}\quad t \gg 1\, .
\end{equation}
By integration last formula over $r$, we simply get
\[
\sqrt {8\pi } \phi_1 \left( {t,r} \right) \approx a\int {\sqrt {\frac{{f'_0  \left( r \right)}}{{rf_0\left( r \right)}}} } dr + \phi _2 \left( t \right)\, ,
\]
for some function $\phi _2 ( t )$ which depends only on time. Let $\int {\sqrt {f_0'  /rf_0} } dr = G( r )$, for some function $G$, so we write
\[
\sqrt {8\pi } \phi_1 \left( {t,r} \right) \approx a G\left( r \right) + \phi _2 \left( t \right)\, .
\]
And by using the condition
\[
 \left. {\phi _1 \left( {t,r} \right)} \right|_{r = R_H }  = 0\, ,
\]
we obtain
\[
\left. {\sqrt {8\pi } \phi _1 \left( {t,r} \right)} \right|_{r = R_H \left( t \right)}  \approx a\left. {G\left( r \right)} \right|_{r = R_H \left( t \right)}  + \phi _2 \left( t \right) = 0\,,
\]
which determines $\phi _2 \left( t \right)$ by
\[
\phi _2 \left( t \right) =  - aG\left( {R_H \left( t \right)} \right) \, .
\]
Therefore, we obtain the solution of $ \phi _1$;
\[
 \sqrt {8\pi }  \phi _1 \left( {t,r} \right) = aG\left( r \right) + \phi _2 \left( t \right) = aG\left( r \right) - aG\left( {R_H \left( t \right)} \right) \, .
\]
By that, we obtain a solution of $\phi ( {t,r} )$ nearby Horizons ($r=R_H$; $K=0$);
\begin{equation}\label{eq:19}
\phi \left( {t,r} \right) = \phi _0 \left( t \right) + \phi _1 \left( {t,r} \right) = \phi _0 \left( t \right) + \frac{aG\left( r \right)}{ \sqrt {8\pi } } - \frac{ aG\left( {R_H \left( t \right)} \right)}{\sqrt {8\pi }} \, .
\end{equation}
But in order to satisfy the equation of $\dot \phi$, equation (\ref{eq:a36}), we need to use the solution (\ref{eq:19}) in that equation. We have
\begin{equation}\label{eq:18}
 \sqrt {8\pi } \dot \phi  \approx b K \sqrt{\frac{{f_0  \left( r \right) f'_0  \left( r \right)  }}{{r}}}     \,  ,\quad \text{at}\quad t \gg 1\, .
\end{equation}
And by using the solution (\ref{eq:19}) in equation (\ref{eq:18}), we obtain 
\[
\dot \phi _0 \left( t \right) - \frac{ a\dot G\left( {R_H \left( t \right)} \right)}{\sqrt {8\pi }} \approx b K \sqrt{\frac{{f_0  \left( r \right) f'_0  \left( r \right) }}{{r}}}  \,.
\]
The independence of the left side of the last equation on $r$ is because of using the approximation $F(t, r) \to f_0 (r) $ in the equation (\ref{eq:a62}). On horizon surfaces, $K= 0$ ($r = R_H (t)$), the last equation becomes
\[
\dot \phi _0 \left( t \right) - \frac{ a\dot G\left( {R_H \left( t \right)} \right)}{\sqrt {8\pi }} = 0 \, ,
\]
or
\begin{equation}\label{eq:21}
 \sqrt {8\pi }  \dot \phi _0 \left( t \right) = a\dot G\left( {R_H \left( t \right)} \right) = \left. {a G'\left( r \right)} \right|_{r = R_H } \dot R_H ,\quad \text{on}\quad r = R_H \,.
\end{equation}
By solving this equation, we obtain the potential function $V(\phi)$ in terms of the function $f_0  ( r )$, but this solution is at late times of the expansion of the universe; regarding the approximation $F(t, r) \to f_0 (r) $, at $t\gg 1$.
\\

From the potential equation (\ref{eq:a41}), we obtain
\[
V\left( {\phi _0 } \right) = \frac{{3H(t)^2 }}{{8\pi }}\, \Rightarrow \phi _0 (t) = V^{ - 1} \left( {\frac{{3H(t)^2 }}{{8\pi }}} \right) = U\left( {\frac{{3H(t)^2 }}{{8\pi }}} \right), \quad \text{for} \quad U = V^{ - 1} \, .
\]
And by taking the derivative of this equation with respect to time, we obtain
\begin{equation}\label{eq:35}
\dot \phi _0 \left( t \right) = U'\left( {\frac{{3H(t)^2 }}{{8\pi }}} \right)\frac{d}{{dt}}\frac{{3H(t)^2 }}{{8\pi }} = U'\left( {\frac{{3H(t)^2 }}{{8\pi }}} \right)\frac{{6H(t)\dot H(t)}}{{8\pi }}  \,.
\end{equation}
Now, if we use the solution (\ref{eq:21}) in equation (\ref{eq:35}), we simply obtain
\[
\frac{1}{ \sqrt {8\pi } }   \left. {a G'\left( r \right)} \right|_{r = R_H } \dot R_H  = U'\left( {\frac{{3H(t)^2 }}{{8\pi }}} \right)\frac{{6H(t)\dot H(t)}}{{8\pi }} \, .
\]
But $G( r )=\int {\sqrt {f'_0  ( r )  /rf_0  ( r )} } dr$, which implies that $G'( r )= {\sqrt {f'_0  ( r ) /rf_0  ( r )} } $, thus we get
\begin{equation}\label{eq:22}
\left. {a\sqrt {\frac{{f'_0   \left( r \right)}}{{rf_0\left( r \right)}}} } \right|_{r = R_H } \dot R_H  = U'\left( {\frac{{3H(t)^2 }}{{8\pi }}} \right)\frac{{6H(t)\dot H(t)}}{\sqrt {8\pi }} \, .
\end{equation}
If the relation between $\dot R_H$ and $\dot H(t)$ is known and unique, the last equation allows us to get a unique relation between the potential $V(\phi)$ and the function $f_0( r )$. 
\\

The horizon surfaces, $r = R_H$, are solutions of $(g_{rr})^{-1}=-K=0$, so we have
\[
(g_{rr})^{-1}=-K =1- \frac{2M(t)}{R_H}  - R_H^2 H(t)^2 =0\, .
\]
For cosmological horizon and at late times of expansion of the universe; $r = R_C\gg M(t)$, so we get $1 - R_C^2 H(t)^2 \approx0$ which implies $R_C H(t) \approx1$. This equation makes a unique relation between $\dot R_C > 0$ and $\dot H(t) <0$, so by using it in equation (\ref{eq:22}), we obtain
\[
\left. {a\sqrt {\frac{{f_0'  \left( r \right)}}{{rf_0\left( r \right)}}} } \right|_{r = R_C } \dot R_C  = U'\left( {\frac{3}{{8\pi R_C^2 }}} \right)\frac{6}{{\sqrt {8\pi } }}\frac{{ - 1}}{{R_C^3 }}\dot R_C \,.
\]
Therefore
\begin{equation}\label{eq:23}
\left. {a\sqrt {\frac{{f_0'  \left( r \right)}}{{rf_0\left( r \right)}}} } \right|_{r = R_C }  = U'\left( {\frac{3}{{8\pi R_C^2 }}} \right)\frac{-6}{{\sqrt {8\pi }R_C^3  }} \,.
\end{equation}
This is a simple relation between the function $f_0( r )$ and the potential $V(\phi)$ of the field $\phi$, where $U=V^{-1}$. Actually the values of $R_C$ do not effects on that relation. The relation (\ref{eq:23}) relates the potential $V(\phi)$ uniquely to the static solution (static part) of the function $F(t,r)$, where at late times of the expansion of the universe ($t>>1$), the function $F(t,r)$ limits to a static function; $f_0( r )$. Oppositely, the relation (\ref{eq:23}) allows us also to obtain the function $f_0( r )$ in terms of a given potential $V(\phi)$, but in this case, may we need to use a formula for the mass function $M(t)$ to shift the identification powers of $t$ in the equation (\ref{eq:a61}).
\\
\\

For example, if we use the solution of the function $f_0( r )$, the equation (\ref{eq:a42});
\[
f_0 (r) = \frac{{c_1 r}}{{\sqrt {r^2  + c_2 } }} \, ,
\]
in the equation (\ref{eq:23}), we get
\[
\frac{{\sqrt {8\pi } R_C^2 }}{{ - 6}}\sqrt {\frac{{c_2 }}{{(R_C^2  + c_2 )}}}  = U'\left( {\frac{3}{{8\pi R_C^2 }}} \right) \, .
\]
As we have suggested before, since the $R_C(t)$, in our solution, increases with respect to the time, it will become $R_C>>c_2$ at late times of the expansion. Therefore, we get 
\[
\frac{{\sqrt {3c_2 } }}{{ - 6}}\sqrt {\frac{{8\pi R_C^2 }}{3}}  = U'\left( {\frac{3}{{8\pi R_C^2 }}} \right)\, ,
\]
which implies that
\[
U'\left( {V_0 } \right) = \frac{{\sqrt {3c_2 } }}{{ - 6}}\frac{1}{{\sqrt {V_0 } }} \,.
\]
Therefore,
\[
U\left( {V_0 } \right) = \frac{{\sqrt {3c_2 } }}{{ - 3}}\sqrt {V_0 }  = V^{ - 1} \left( {V_0 } \right) \Rightarrow V\left( {\frac{{\sqrt {3c_2 } }}{{ - 3}}\sqrt {V_0 } } \right) = V_0  \, ,
\]
which has the solution
\[
V\left( \phi  \right) = \frac{9}{{3c_2 }}\phi ^2  = \frac{3}{{c_2 }}\phi ^2 \,.
\]
By that, we have obtained the approximated solution for the potential function $V( \phi  )$ of the field $\phi$, at late times of the expansion. It is more convenient to choose $c_2=6$, to get $V( \phi  ) = \phi ^2/2$. Note that this potential associates with a certain solution of the function $F(t,r)$, the solution (\ref{eq:a33}).

\section{Conclusion}
We have tested some solutions of the spherical symmetric gravitational interaction of a black hole with a real scalar field in background of expanding universe by using some suitable metric. And we have obtained the solutions of the field $\phi$, the black hole mass $M(t)>0$ and the expansion rate $H(t)>0$. We suggested two behaviours for decreasing of the universal energy density during the  expansion of the universe, on with powers of $1/t$ (slow expansion), and other with powers of $\exp(-t)$ (fast expansion). In both solutions, we found that the mass of the black hole is stable and remains constant at all late times of the cosmological expansion, and the solutions of the field $\phi$ are concentrated (or created) around $r=0$ inside the black hole (as seen by exterior and interior observers), we relate that behaviour to the gravity of the black hole. We found that the energy density of the field $\phi(t,r)$ inside of the black hole (as measured by some interior observer) is almost stable under the influence of the cosmological expansion, by the meaning that it tends to take non-vanishing values. In the second solution, solution of $\exp(-t)$-power series, the field $\phi$ can take a constant value at late times of the expansion, and so obtaining an accelerated cosmological expansion. We also found that there is a relation (equation) that relates the potential $V(\phi)$ uniquely to the static solution of the function $F(t,r)$; the solution $f_0(r)$. 
\\

{\bf Funding and/or Conflicts of interests/Competing interests:}

There is no funding and/or conflicts of interests/competing interests regarding this manuscript.\\

{\bf Data Availability Statement}:

 No Data associated in the manuscript.

\section{Appendix}
In this appendix, we derive the differential equation for the function $F(t,r)$, and find that the root term 
\[
\sqrt {\left( {\frac{{F^\prime  }}{{rF}}} \right)^2  - \frac{{4W^2 }}{{r^4 K^4 F^2 }}}
\]
indeed vanishes.
\\

From the potential equation (\ref{eq:17});
\begin{equation}
8\pi \left. {V\left( \phi  \right)} \right|_{\phi \left( {t,r} \right)}  = 3H(t)^2  + \frac{{F'(t,r)}}{{rF(t,r)}}K\,,
\end{equation}
and by considering $\phi ( {t,r} )$ as a solution of the equations, we obtain (by taking the derivative with respect to $t$ and $r$);
\begin{equation}\label{eq:a46}
\begin{split}
& 8\pi \left. {V'\left( \phi  \right)} \right|_{\phi \left( {t,r} \right)} \phi '\left( {t,r} \right) = \left( {AK} \right)' = d\,, \quad \text{for}\quad  A= \frac{{F'(t,r)}}{{rF(t,r)}}\, ,\\ 
&\\
& 8\pi \left. {V'\left( \phi  \right)} \right|_{\phi \left( {t,r} \right)} \dot \phi \left( {t,r} \right) = 6H(t)\dot H(t) + \dot AK + A\dot K\, , \\
&\\
\end{split}
\end{equation}
for $V'\left( \phi  \right) = dV\left( \phi  \right)/d\phi $. And from
\[
W = r^3 H\dot H + \dot M,\quad \text{where}\quad \dot K = \frac{2}{r}W\, ,
\]
we get
\[
H\dot H = \frac{{W - \dot M}}{{r^3 }},
\]
by using it in the second equation of (\ref{eq:a46}), we get
\[
8\pi \left. {V'\left( \phi  \right)} \right|_{\phi \left( {t,r} \right)} \dot \phi \left( {t,r} \right) = 6\frac{{W - \dot M}}{{r^3 }} + \dot AK + A\frac{2}{r}W = b + cW \,  ,
\]
for
\[
b = \dot AK - \frac{{6\dot M}}{{r^3 }},\quad \text{and}\quad  c = \frac{6}{{r^3 }} + \frac{2}{r}A \, .
\]
By that, the equations (\ref{eq:a46}) give
\begin{equation}\label{eq:a47}
\frac{{8\pi \left. {V'\left( \phi  \right)} \right|_{\phi \left( {t,r} \right)} \dot \phi \left( {t,r} \right)}}{{8\pi \left. {V'\left( \phi  \right)} \right|_{\phi \left( {t,r} \right)} \phi '\left( {t,r} \right)}} = \frac{{\dot \phi \left( {t,r} \right)}}{{\phi '\left( {t,r} \right)}} = \frac{{b + cW}}{d} \, .
\end{equation}
In other hand, from the equations ((\ref{eq:6}) and (\ref{eq:14}));
\begin{equation}\label{eq:a52}
\begin{split}
&8\pi \phi {^\prime}  ^2  = \frac{{F^\prime  }}{{rF}} + a\sqrt {\left( {\frac{{F^\prime  }}{{rF}}} \right)^2  - \frac{{4W^2 }}{{r^4 K^4 F^2 }}} \,,\\
&\\
&8\pi \dot \phi ^2  = \frac{{FF^\prime  }}{r}K^2  - aK^2 F^2 \sqrt {\left( {\frac{{F^\prime  }}{{rF}}} \right)^2  - \frac{{4W^2 }}{{r^4 K^4 F^2 }}} \,,
\end{split}
\end{equation}
we obtain 
\[
\frac{{\dot \phi ^2 }}{{\phi {^\prime}  ^2 }} = \frac{{\frac{{2FF^\prime  }}{r}K^2 }}{{A + \sqrt {A^2  - B^2 } }} - K^2 F^2 ,\quad \text{for}\quad B = eW = \frac{2}{{r^2 K^2 F}}W\, .
\]
Using this equation in equation (\ref{eq:a47}), we get
\[
\frac{{\left( {b + cW} \right)^2 }}{{d^2 }} = \frac{{\frac{{2FF^\prime  }}{r}K^2 }}{{A + \sqrt {A^2  - B^2 } }} - a^2 ,\quad \text{for}\quad a = KF \, .
\]
After isolating the root term and taking the square, we get
\[
4A^2 a^2 \frac{{\left( {b + cW} \right)^2 }}{{d^2 }} = \left( {a^2  + \frac{{\left( {b + cW} \right)^2 }}{{d^2 }}} \right)^2 e^2 W^2 \, ,
\]
therefore
\begin{equation}\label{eq:a50}
2Aa\frac{{\left( {b + cW} \right)}}{d} =-  c_1 \left( {a^2  + \frac{{\left( {b + cW} \right)^2 }}{{d^2 }}} \right)eW \,, \quad \text{where}\quad  c_1=\pm 1\, ,
\end{equation}
so
\[
eW\frac{{\left( {b + cW} \right)^2 }}{{d^2 }} +c_1 2Aa\frac{{\left( {b + cW} \right)}}{d} + a^2 eW = 0 \, .
\]
This equation has the solution
\[
\frac{{b + cW}}{d} = \frac{{ -  c_1 2Aa \mp \sqrt {\left( {2Aa} \right)^2  - 4\left( {eW} \right)\left( {a^2 eW} \right)} }}{{2eW}}\, .
\]
Or
\[
\frac{{b + cW}}{d} = \frac{a}{{eW}}\left( { -  c_1 A \mp \sqrt {A^2  - e^2 W^2 } } \right) \, .
\]
If we choose the solution $  c_1=-1$:
\begin{equation}\label{eq:a49}
\frac{{b + cW}}{d} = \frac{a}{{eW}}\left( {A \mp \sqrt {A^2  - e^2 W^2 } } \right) \, ,
\end{equation}
we obtain 
\begin{equation}
\frac{{b + cW}}{d}\frac{{eW}}{a} = A \mp \sqrt {A^2  - e^2 W^2 }   \, .
\end{equation}
Comparing with equation of $\phi {'} $, first equation of (\ref{eq:a52}), we obtain another formula for $\phi {'}$;
\begin{equation}\label{eq:a48}
8\pi \phi {'}  ^2=\frac{{b + cW}}{d}\frac{{eW}}{a}   \, .
\end{equation}
And by using this formula in second equation of (\ref{eq:3});
\[
G_{tr}  =  - \frac{{2\left( {r^3 H(t)\dot H(t) + \dot M(t)} \right)}}{{r\left( {r^3 H(t)^2  + 2M(t) - r} \right)}} =  - \frac{{2W}}{{r^2 K}} = 8\pi \dot \phi \phi '\,,
\]
we also obtain another equation for $ \dot \phi$;
\[
8\pi \dot \phi ^2  = \frac{{2F^2 Kd}}{{r^2 }}\frac{W}{{b + cW}} \, .
\]
And by using this equation with equation (\ref{eq:a48}) in the equation (\ref{eq:25}), we get
\[
\frac{{2FF^\prime  }}{r}K^2=\frac{{2F^2 Kd}}{{r^2 }}\frac{W}{{b + cW}} +K^2 F^2 \frac{{\left( {b + cW} \right)eW}}{{ad}}    \, ,
\]
therefore
\[
\frac{{2F^2 Kd}}{{r^2 }}\left( {\frac{W}{{b + cW}}} \right)^2  - \frac{{2FF^\prime  K^2 }}{r}\left( {\frac{W}{{b + cW}}} \right) + \frac{{K^2 F^2 eW^2 }}{{ad}} = 0 \, ,
\]
which has the solution
\[
\frac{W}{{b + cW}} = \frac{{\frac{{2FF^\prime  K^2 }}{r} \mp \sqrt {\left( {\frac{{2FF^\prime  K^2 }}{r}} \right)^2  - 4\left( {\frac{{2F^2 Kd}}{{r^2 }}} \right)\left( {\frac{{K^2 F^2 eW^2 }}{{ad}}} \right)} }}{{4\frac{{2F^2 Kd}}{{r^2 }}}} \, ,
\]
or
\[
\frac{W}{{b + cW}} = \frac{{r^2 K}}{{2d}}\left( {A \mp \sqrt {A^2  - e^2 W^2 } } \right) \, .
\]
And by using this equation in (\ref{eq:a49}), we get
\[
b + cW =  \mp KFd \, .
\]
And by using this equation in (\ref{eq:a50}), we finally get
\begin{equation}\label{eq:a51}
\begin{split}
 &\left( {a^2  + K^2 F^2 } \right)eW =  + 2Aa(\mp 1)KF\, , \\ 
&\\ 
& \Rightarrow eW =  \mp A \, , \quad \text{so} \quad W =\mp \frac{1}{2}rK^2 F' \, .
\end{split}
\end{equation}
We choose '$+$' sign solution, because as we think, $W$ is mostly negative (for $\dot H<0$) and we expect that $F'<0$, therefore
\begin{equation}\label{eq:a26}
W = \frac{1}{2}rK^2 F' \, .
\end{equation}
This solution of $W$ vanishes the root term in the equations (\ref{eq:a52}); we obtain
\begin{equation}\label{eq:a53}
\begin{split}
& 8\pi \phi {^\prime}  ^2  = \frac{{F^\prime  }}{{rF}}\,, \quad \text{and} \quad 8\pi \dot \phi ^2  = \frac{{FF^\prime  }}{r}K^2 \,, \\ 
  \Rightarrow  \\ 
& \sqrt {8\pi } \phi ^\prime   = a\sqrt {\frac{{F^\prime  }}{{rF}}} \,,\quad \text{and}\quad \sqrt {8\pi } \dot \phi   = b\sqrt {\frac{{FF^\prime  }}{r}} \, K \,,  
\end{split}
\end{equation}
for $a,b=\mp 1$. 
\\

These solutions imply
\[
\sqrt {8\pi } \partial _0 \phi ^\prime   = a\frac{{\dot A}}{{2\sqrt A }}\,,\quad \text{and}\quad \sqrt {8\pi } \partial _r \dot \phi  = b\frac{{B'}}{{2\sqrt B }}K + b\sqrt B K' \, ,
\]
where
\[
A = \frac{{F^\prime  }}{{rF}} \,,\quad \text{and}\quad B = \frac{{FF^\prime  }}{r} \, .
\]
Therefore, the consistence condition, $  \partial _t \phi ^\prime=\partial _r \dot \phi  $, yields
\[
a\frac{{\dot A}}{{2\sqrt A }} = b\frac{{B'}}{{2\sqrt B }}K + b\sqrt B K' \, .
\]
Using
\[
K' = \frac{2}{r}K + \frac{2}{r}\left( {1 - \frac{{3M\left( t \right)}}{r}} \right) \, ,
\]
we obtain
\[
a\frac{{\dot A}}{{2\sqrt A }} = b\left( {\frac{{B'}}{{2\sqrt B }} + \frac{2}{r}\sqrt B } \right)K + b\frac{2}{r}\left( {1 - \frac{{3M\left( t \right)}}{r}} \right)\sqrt B  \, .
\]
Which becomes
\begin{equation}\label{eq:a25}
\frac{{ab}}{2}F\dot A - \frac{{2B}}{r}\left( {1 - \frac{{3M\left( t \right)}}{r}} \right) = \left( {\frac{{B'}}{2} + \frac{2}{r}B} \right)K \, .
\end{equation}
We also have the equation (\ref{eq:10}), with vanishing the root term, it becomes
\[
 - \frac{{\,F''}}{F} - 3\left( {1 + \frac{{r - 3M}}{{rK}}} \right)\frac{{F'}}{{rF}} - \frac{1}{{r^2 K^4 F^2 }}\left( {rK\dot W - 4W^2 } \right) + \frac{{\dot FW}}{{rK^3 F^3 }}\, = 0 \, ,
\]
and by using the solution (\ref{eq:a26}), it becomes
\begin{equation}\label{eq:a27}
2\left( { - \,FF'' - \frac{{3FF'}}{r}} \right)KF - 6\left( {\frac{{r - 3M}}{r}} \right)\frac{{F^2 F'}}{r} + \dot FF' - F\dot F'\, = 0 \, .
\end{equation}
Omitting $K$ from the equations (\ref{eq:a25}) and (\ref{eq:a27}), we get
\begin{equation}
\begin{split}
& 2\left( { - \,FF'' - \frac{{3FF'}}{r}} \right)F^2 \left[ {\frac{{ab}}{2}F\dot A - \frac{{2B}}{r}\left( {1 - \frac{{3M\left( t \right)}}{r}} \right)} \right] \\ 
&\\
&  + \left[ { - 6\left( {\frac{{r - 3M\left( t \right)}}{r}} \right)\frac{{F^2 F'}}{r} + \dot FF' - F\dot F'\,} \right]\left( {\frac{{B'}}{2} + \frac{2}{r}B} \right) = 0 \, . 
\end{split}
\end{equation}
Using $A={F'/rF}$ and $B={FF'/r}$, this equation becomes
\begin{equation}\label{eq:a28}
\begin{split}
& \left( {F\dot F' - \dot FF'} \right)\left[ {F'^2  + \left( {\,FF'' + \frac{{3FF'}}{r}} \right)\left( {2ab + 1} \right)} \right] \\ 
&\\
&  - \frac{{2F^2 F'}}{{r^2 }}\left( {1 - \frac{{3M\left( t \right)}}{r}} \right)\left[ { - 3F'^2  + \,FF'' + \frac{{3FF'}}{r}} \right] = 0 \, . 
\end{split}
\end{equation}
This is a differential equation for the function $F(t,r)$. But it includes the mass function $M(t)$. However, the black hole is almost stable, therefore we can approximate $M(t)\approx M_0$, for some constant mass $M_0$. By solving this equation and obtaining the function $F(t,r)$, we can obtain the other functions in term of it.

\end{document}